# Tunable spin and orbital torques in Cu-based magnetic heterostructures

Silvia Damerio* and Can O. Avci

*Institut de Ciència de Materials de Barcelona, Campus de la UAB, Bellaterra, 08193, Spain*

E-mail: sdamerio@icmab.es

**Abstract**

Current-induced torques originating from earth-abundant $3d$ elements offer a promising avenue for low-cost and sustainable spintronic memory and logic applications. Recently, orbital currents -transverse orbital angular momentum flow in response to an electric field- have been in the spotlight since they allow current-induced torque generation from $3d$ transition metals. Here, we report a comprehensive study of the current-induced spin and orbital torques in Cu-based magnetic heterostructures. We show that high torque efficiencies can be achieved in engineered $Ni_{80}Fe_{20}$/Cu bilayers where Cu is naturally oxidized, exceeding the ones found in the archetypical Co/Pt. Furthermore, we demonstrate sign and amplitude control of the damping-like torque by manipulating the oxidation state of Cu via solid-state gating. Our findings provide insights into the interplay between charge, spin, and orbital transport in Cu-based heterostructures and open the door to the development of gate-tunable spin-orbitronic devices.

The interconversion between charge and spin currents is a central topic in spintronics, allowing magnetization control through spin-orbit torques (SOTs).[1,2] So far, the generation of SOTs has mostly relied on the spin-Hall (SHE) and interfacial Rashba-Edelstein (REE) effects in materials characterized by strong spin-orbit coupling, thus limiting the choice of suitable materials to heavy metals such as Pt, Ta, and W. Recently, the orbital counterpart of the SHE, the orbital Hall effect (OHE), which consists in the flow of orbital angular momentum perpendicular to a charge current, has drawn attention as an alternative route for the generation of orbital torques (OTs).[3–5] Theory suggests that OHE is more common and fundamental than the SHE as it can occur in any material without requiring a strong spin-orbit coupling.[6] In several $3d$ transition metals, the orbital Hall conductivity has been predicted to largely exceed the spin Hall conductivity of the best-known materials.[7] Consequently, a significant effort has been recently devoted to the demonstration of orbital currents, orbital accumulation, and its conversion to orbital torques in ferromagnetic heterostructures.[8–15]

Cu is one of the most intriguing materials in the OHE context. Earlier reports showed that the Cu/oxide interface can act as a spin-torque generator.[16,17] Multiple studies followed on the effects of natural oxidation[18–23] and nitration[24] of Cu, as well as its interface engineering with adjacent oxide materials[25,26] for the enhancement of the OT efficiency. Other effects such as unidirectional orbital magnetoresistance,[27] orbital Hanle magnetoresistance,[28] and orbital REE[11,29] were discovered in Cu-based heterostructures underlining an intricate relation between charge, orbital and spin degrees of freedom in this material. However, despite extensive studies, fundamental questions remain regarding the nature of the orbital texture required to observe the OHE in Cu. In particular, clear distinctions between spin vs. orbital and bulk vs. interface contributions are far from established. Because orbital accumulation cannot directly interact with magnetization due to the lack of exchange coupling between orbital angular momentum and local magnetization, orbital-to-spin conversion must occur either at the interface or in the bulk of the adjacent ferromagnet (FM).[30] This adds yet another complexity to understanding the underlying physics and finding the most suitable material combinations, which is essential to designing new device architectures for efficient spin-orbitronics applications.

In this study, we examined the (S)OTs induced in Co and $Ni_{80}Fe_{20}$ (Py) by adjacent layers of Cu, Pt and naturally oxidized $CuO_x$ layers. In structures with $CuO_x$, we observe a damping-like OT with the opposite sign to the Pt and Cu references. Additionally, the FM thickness dependence with a fixed CuOx layer demonstrates that the damping-like OT component increases significantly with increasing thickness of Co and Py. In particular in Py / $CuO_x$, the magnitude of OT efficiency becomes comparable to the SOT generated in the reference Py/Pt bilayers and exceeds that of Co/Pt when the Py thickness reaches 5 nm. A collective understanding of the data pinpoints the OHE origin of the damping-like OT in FM/$CuO_x$ structures. In addition, we demonstrate full electrical tunability of the damping-like OT in Co/$CuO_x$/$GdO_x$ devices by solid-state gating. This is achieved by precisely and reversibly controlling the oxidation state of the $CuO_x$ layer by voltage-driven ion migration. Our findings support the potential for using Cu to achieve efficient electrical control of magnetization, thereby encouraging further exploration of spin-orbitronic devices based on more economical and sustainable light elements.

Ti(1.5)/FM($t_{FM}$)/NM(3) (thicknesses in nm) with easy-plane magnetic anisotropy were deposited by d.c. magnetron sputtering onto $SiO_x$ substrates (methods in Supporting Information Note 1). Here, Ti was used to improve adhesion and plays no active role in the device functionality; hence, it will not be mentioned further. The FMs are Co or Py, whereas the NMs are Cu or Pt. To investigate the effect of the natural oxidation of Cu on the (S)OTs, we either capped the final structure with 1.5 nm of Ti or left it uncapped to obtain unoxidized and partially oxidized Cu, respectively (see Supporting Information Note 2). $t_{FM}$ was varied between 1.5 and 10 nm for each FM/NM combination. The saturation magnetization ($M_s$) of the Co and Py was measured for different $t_{FM}$ and found between 1.17 - 1.25 MA/m for Co depending on the thickness, and 0.68 MA/m for Py, independently of its thickness (see Supporting Information Note 3). These values agree with previous reports[12,19] and indicate good quality of the deposited FM films. To quantify the current-induced (S)OTs, we patterned the heterostructures in Hall bars, depicted in Figure 1a, and performed harmonic Hall measurements.[31,32]

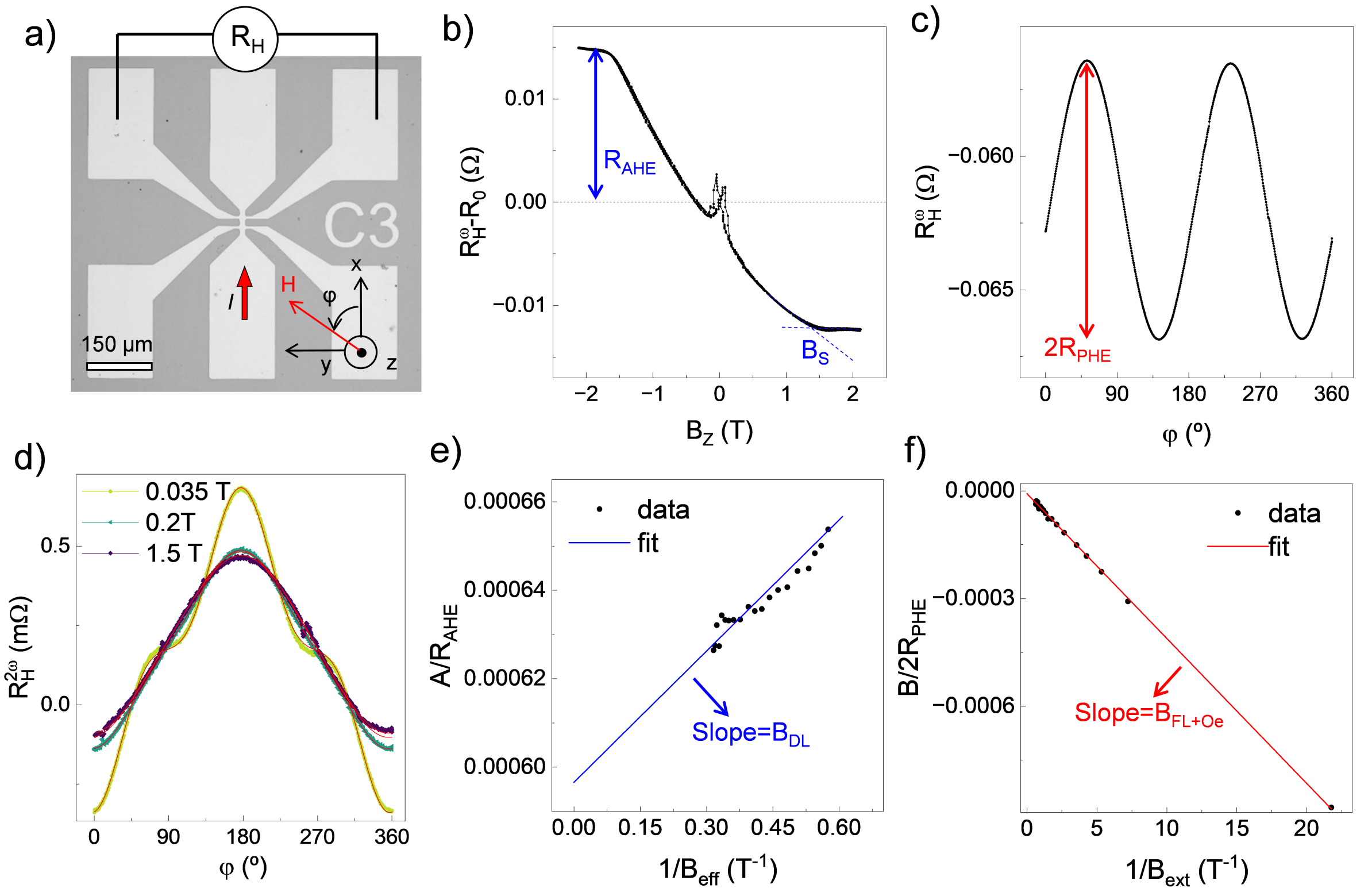

Figure 1: SOT quantification by harmonic Hall measurements. a) Optical microscopy image of a representative Hall bar device and measurement scheme. b) First harmonic Hall resistance ($R_H^{\omega}$) as a function of out-of-plane field ($B_Z$). c) $R_H^{\omega}$ as a function of in-plane angle ($\varphi$) measured at 1T. d) Second harmonic Hall resistance ($R_H^{2\omega}$) as a function $\varphi$ measured at three representative external fields from 0.035 to 1.5 T. e) Plot of the normalized $cos\varphi$ component of $R_H^{2\omega}(\varphi)$ as a function of inverse effective field ($1/B_{eff}$) and f) plot of the normalized $2cos^3\varphi - cos\varphi$ component of $R_H^{2\omega}(\varphi)$ as a function of inverse external field ($1/B_{ext}$).

First, we measure the first-harmonic transverse resistance ($R_H^{\omega}$) as a function of a swept out-of-plane magnetic field ($B_Z$) and extract the anomalous Hall resistance ($R_{AHE}$) and saturation field ($B_S$), as shown in Figure 1b. Then, we record $R_H^{\omega}$ and the second harmonic response ($R_H^{2\omega}$) as a function of in-plane angle ($\varphi$) with constant external field ($B_{ext}$). The former provides the planar Hall resistance ($R_{PHE}$), as shown in Figure 1c, while the latter, plotted in Figure 1d for 3 representative fields, contains the information on the current induced SOTs by reflecting the motion of the magnetization vector under the applied a.c. current ($I_\omega$). $R_H^{2\omega}$ is given by:[30,33]

$$R_H^{2\omega}(\varphi) = [R_{AHE}(B_{DL}/B_{eff}) + R_{\Delta T}]cos\varphi + [2R_{PHE}(B_{FL} + B_{eff})/B_{ext}](2cos^3\varphi - cos\varphi) \quad (1)$$

where $B_{DL}$ and $B_{FL}$ are the damping-like and field-like SOT effective fields; $B_{Oe}$ is the current-induced Oersted field; $B_{eff}$ is the effective magnetic field calculated as $B_{eff} = B_{ext} + B_S$ and $R_{\Delta T}$ represents the thermal contributions predominantly from the anomalous and spin Seebeck effects due to an out-of-plane temperature gradient as described in Ref.[32]. Therefore, the damping-like and field-like (S)OT contributions can be separated based on their angular dependence from the fit of $R_H^{2\omega}(\varphi)$ with a function of the type $Acos\varphi + B(2cos^3\varphi - cos\varphi)$ (red in Figure 1d). $B_{DL}$ and $B_{FL}$ are thus obtained from the slope of the normalized $cos\varphi$ and $2cos^3\varphi - cos\varphi$ components as a function of $1/B_{eff}$ and $1/B_{ext}$, respectively (Figure 1e-f):

$$A/R_{AHE} = B_{DL}/B_{eff} \quad (2)$$

$$B/2R_{PHE} = (B_{FL} + B_{Oe})/B_{ext} \quad (3)$$

Note that $B_{Oe}$ contribution can be estimated by the current in the NM layer ($I_{NM}$) using a parallel resistor model (see Supporting Information Note 4) and following the formula $B_{Oe} = \mu_0 I_{NM}/2 * w$,[34,35] where $w$ =7.5 µm is the width of the Hall bar. Finally, the torque efficiency is calculated taking into account $M_S$ and $t_{FM}$ and normalized by the applied electric field ($E$):

$$\xi^E_{SOT} = (2e/\hbar)B_{SOT}t_{FM}M_S/E \quad (4)$$

(S)OTs quantification for Co($t$)/NM(3) bilayers is summarized in Figure 2. Figure 2a shows the values of $B_{DL}$ normalized by $E$ as a function of $t_{Co}$ in Co($t$)/CuO$_x$ (red penthagon), Co($t$)/Cu (orange triangle) and Co($t$)/Pt (gray circle). In a typical SOT system such as Co/Pt where $B_{DL}$ is predominantly driven by the SHE, $B_{DL}$ is expected to decrease with the increasing $t_{Co}$,[22] as is the case in our Co/Pt data. However, in Co/CuO$_x$ and Co/Cu $B_{DL}$ increases with $t_{Co}$ instead. Here, $B_{DL}$ is more than an order of magnitude lower than those generated from Pt, indicating generally low efficiency of using Cu as the torque generator acting upon Co. However, when naturally oxidized CuO$_x$ is used, $B_{DL}$ has the opposite (negative) sign compared to the Co/Cu and reference Co/Pt bilayers, but the same sign as the inverted stack Pt/Co (squares in Figure 2a-b). This intriguing result will be elaborated on later. Figure 2b shows the corresponding damping-like (S)OT efficiency ($\xi^E_{DL}$) calculated following eq.(4). For Co/Pt, the magnitude of $\xi^E_{DL}$ is in line with previous reports[1] and remains essentially constant (gray line in Figure 2b) with a minor increase for $t_{Co} > 5$ nm. On the other hand, $\xi^E_{DL}$ increases monotonically in Co/CuO$_x$ and Co/Cu as a function of $t_{Co}$ (note that the values are multiplied by 4 for better readability), indicating an enhanced efficiency with an increased magnetic volume, contrary to the well-established understanding of conventional SOTs.

Now, we turn to the field-like (S)OT in Co($t$)/NM(3) structures. Figure 2c shows $B_{FL}$ as a function of $t_{Co}$ after subtracting the $B_{Oe}$ contribution from the measured quantity following a parallel resistor model (see Supporting Information Note 4). In Co/Pt, $B_{FL}$ is negative and decreases in magnitude with the increasing $t_{Co}$, following approximately the same trend as $B_{DL}$, pinpointing their common origin. On the other hand, $B_{FL}$ is positive in both Co/Cu and Co/CuO$_x$ with comparable magnitude to those obtained in Co/Pt. Both datasets show a similar decreasing trend as a function of $t_{Co}$, contrasting with the $B_{DL}$ trends observed in these systems. This observation indicates a different origin of $B_{FL}$ than that responsible for $B_{DL}$ in Co/Cu and Co/CuO$_x$. We tentatively attribute the source of $B_{FL}$ to the Co/Cu interface, common in these two systems, where mechanisms like the interfacial REE could be at play. Overall, the SOTs quantification in Figure 2 suggests that Cu, $CuO_x$, and their interfaces with Co are potential sources of (S)OTs. The opposite sign of $B_{DL}$ in Cu vs. $CuO_x$ and their $t_{Co}$ dependence provide strong evidence of an unconventional torque component that could be attributed to an orbital angular momentum accumulation in $CuO_x$ that is then converted to spin-torque in Co.

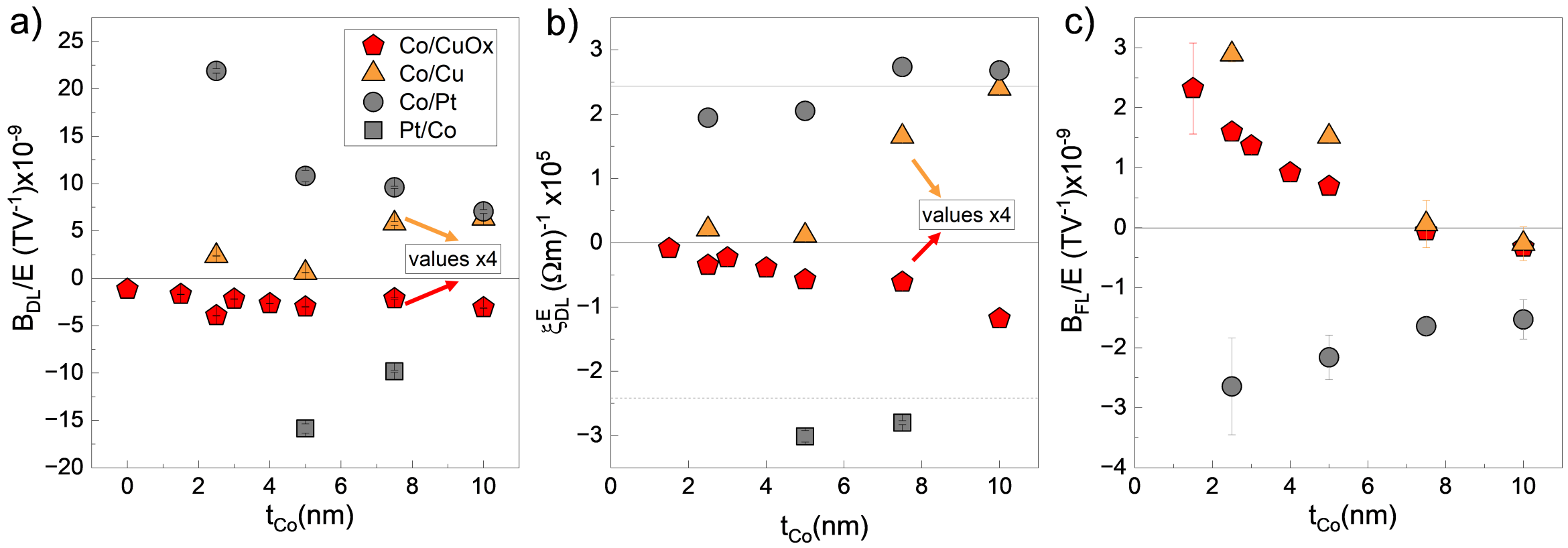

Figure 2: (S)OTs in Co($t$)/NM(3) bilayers. Plot of a) the damping-like torque normalized effective field ($B_{DL}/E$), b) the damping-like torque efficiency ($\xi^E_{DL}$) and c) the field-like torque normalized effective field ($B_{FL}/E$) as a function of Co thickness ($t_{Co}$) for NM=$CuO_x$ (red pentagon), NM=Cu (orange triangle), NM=Pt (gray circle and square).

Next, we measured (S)OTs in equivalent Py($t$)/NM structures to test the above hypothesis. Here, replacing Co with Py allows better harnessing of the OHE contribution to the torques, given the larger spin-orbit conversion coefficient of Py.[9,36] The (S)OT results for Py($t$)/NM bilayers are summarized in Figure 3. Similarly to the Co-based bilayers, here $B_{DL}$ also scales with $1/t_{Py}$ when Pt is used as torque generator (gray circles in Figure 3a), and changes sign (but not in amplitude) when the stack is inverted (gray squares), establishing the SHE as the main source for $B_{DL}$. In contrast, in Py/$CuO_x$ (red pentagon in Figure 3a), $B_{DL}$ increases as a function of $t_{Py}$ up to 5 nm and then follows a decreasing trend. Similar behavior was observed in CoFe/$CuO_x$[22] and associated with the orbital origin of the spin accumulation in $CuO_x$. For the orbital accumulation to generate torque on the magnetization, the orbital-to-spin conversion needs to occur in the first few nm of the FM layer. This leads to an initial sharp increase of $B_{DL}$ with $t_{FM}$ followed by an exponential decrease as the process saturates. Finally, in Py/Cu (orange triangle in Figure 3a), a very low and positive SOT is observed.

Figure 3b shows $\xi^{E}_{DL}$ for all Py($t$)/NM structures with a highly contrasting outcome to that found in Co($t$)/NM(3) reported in Figure 2b. First, $\xi^{E}_{DL}$ in Py/CuO$_x$ is significantly larger than in Co/Pt, Co/CuO$_x$ and Py/Cu, approaching the efficiency found in Py/Pt (gray dashed line). Second, $\xi^{E}_{DL}$ of Py/Pt is constant up to 5 nm with magnitudes up to three times larger than that found in Co/Pt and increases further for larger $t_{Py}$. The significantly enhanced SOT efficiency compared to Co/Pt, together with the increasing trend for larger Py thickness, could indicate OHE contribution in Pt, which is reported to have the same sign as the SHE.[7] Another possibility is the self-induced torque in Py.[37] However, we characterized this in single Co and Py layers, and $B_{DL}$ was found to be negligible in both cases (see Supporting Information Note 5). Conversely, with CuO$_x$, $\xi^{E}_{DL}$ sharply increases up to 5 nm and then tends to saturate to a value comparable to the $\xi^{E}_{DL}$ of Pt, but with the opposite sign (dotted gray line in Figure 3b). As mentioned earlier, this trend corroborates with the expected behavior of orbital currents, which need a certain amount of magnetic thickness to be fully converted into a spin current and hence exert torque on the magnetization. This thickness is characterized as the orbital diffusion length and depends on the FM choice. Assuming that $B_{DL}$ is predominantly driven by orbital currents, in the case of Py, the orbital diffusion length is around 5 nm, whereas in Co, it seems to be >10 nm (see Supporting Information Note 6). However, further experiments are required to validate these numbers and assumptions.

Finally, the $t_{Py}$ dependence of $B_{FL}$ (after subtracting the $B_{Oe}$ contribution) is displayed in Figure 3c. Here, contrary to the situation with $B_{DL}$, $B_{FL}$ shows similarity to the case of Co($t$)/NM. $B_{FL}$ in Py/Pt is slightly larger than in Co/Pt, although the error bar is also large. $B_{FL}$ in Py/Cu and Py/CuO$_x$ shows the same qualitative trend of Co-based structures but it undergoes a sign reversal and saturation towards a small negative value.

We do not have a plausible explanation for the sign crossover and non-negligible $B_{FL}$ at large Py thicknesses in Cu and $CuO_x$ samples except the possible underestimation of the $B_{Oe}$ contribution, which would erroneously lead to the reported trend.

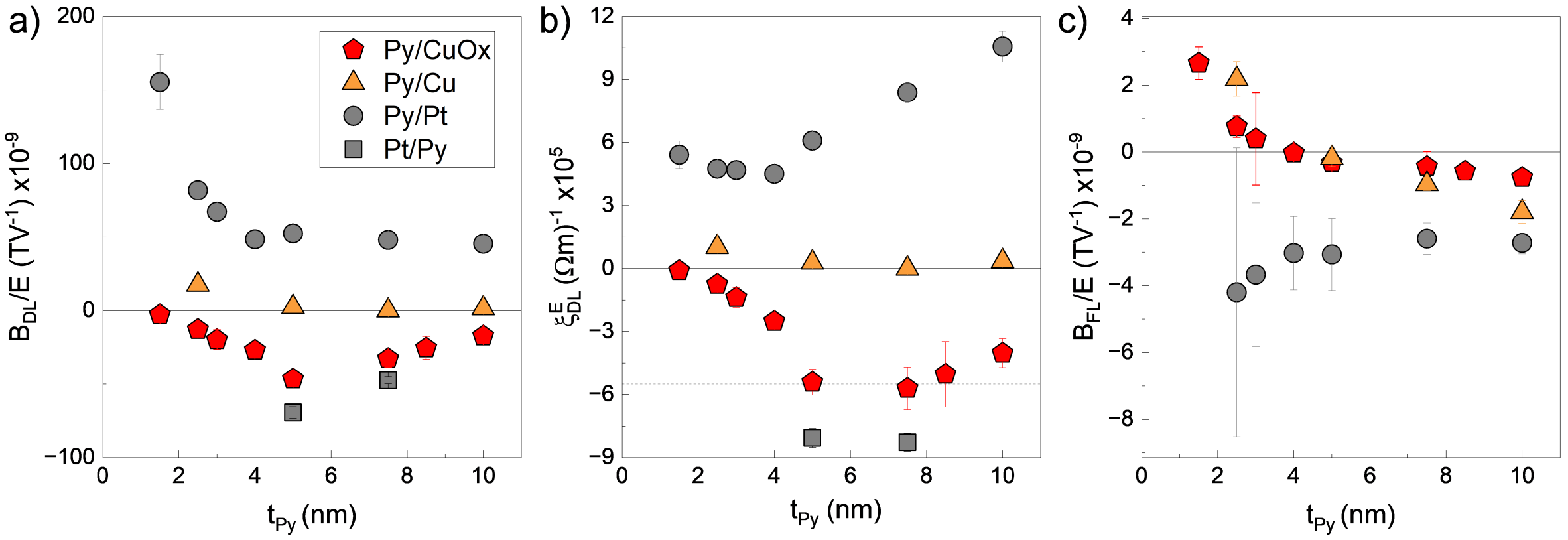

Figure 3: SOTs in Py(t)/NM(3) bilayers. Plot of a) the damping-like torque normalized effective field ($B_{DL}/E$), b) the damping-like torque efficiency ($\xi_{DL}^{E}$) and c) the field-like torque normalized effective field ($B_{FL}/E$) as a function of Py thickness ($t_{Py}$) for NM=$CuO_x$ (red pentagon), NM=Cu (orange triangle), NM=Pt (gray circle and square).

The large difference of $\xi_{DL}^{E}$ between Co/$CuO_x$ and Py/$CuO_x$ can be explained by considering the spin-orbit conversion mechanism at the FM/$CuO_x$ interfaces. In FM/NM bilayers, the overall charge-to-spin conversion efficiency ($\theta_{SH}$) is given by the sum of spin ($\sigma_{SH}^{NM}$) and orbital ($\sigma_{OH}^{NM}$) conductivity normalized by the electrical conductivity ($\sigma^{NM}$):[9]

$$\theta_{SH} = (2e/\hbar)(\sigma_{SH}^{NM} + \sigma_{OH}^{NM}\eta_{L-S}^{FM})/\sigma^{NM} \tag{5}$$

As apparent from eq.(5), the orbital contribution to $\theta_{SH}$ depends on the spin-orbit conversion coefficient ($\eta_{L-S}^{FM}$) of the ferromagnet. While for Py, due to its high Ni content, we expect a large spin-orbit conversion coefficient, this term is much smaller in Co ($\eta_{L-S}^{Ni}>\eta_{L-S}^{Co}$).[36] Hence, the orbital moment accumulation generated due to the OHE in $CuO_x$ is more efficiently converted in a torque on the magnetization of Py than Co. Moreover, when Cu is not oxidized, we observe a small positive amplitude of the damping-like torque in Co/Cu bilayers and almost no torque in Py/Cu. This is in agreement with theoretical calculations[8] that predict both $\sigma_{SH}^{NM}$ and $\sigma_{OH}^{NM}$ to be small and positive in metallic Cu.[8] It is plausible to assume that in the Py/Cu case, a small reminiscent oxidation counteracts the spin hall effect, making the total torque even smaller than in the Co/Cu case. Nevertheless, we find that natural oxidation of Cu is a viable method to increase the generation of orbital currents in this light metal.

The negative sign of $\xi_{DL}^{E}$ in the FM/$CuO_x$ bilayers of this study and its dependence on $t_{FM}$ both point towards an orbital origin of the spin accumulation. Negative-sign torque on Co has also been observed before in Co/$CuN_x$ by Chen *et al.*,[24] and its sign had been attributed to a negative L-S conversion coefficient ($\eta_{L-S}^{Co}$) of Co, whereas An *et al.* reported positive orbital torques in Py/$CuO_x$.[16] On the other hand, Xiao et al.[19] report a significant enhancement of the OT efficiency in Pt/Co/Cu–$CuO_x$ heterostructures, where the OT derived from orbital angular momentum accumulation in Cu-$CuO_x$ and the SOT provided by the Pt layer drive the rotation of the magnetization collaboratively. Considered the geometry of the stack, this means that the OT from Cu-$CuO_x$ and the SOT from Pt on Co have opposite sign, which corresponds to our observations. These discrepancies denote that the mechanism for the orbital angular momentum accumulation is highly dependent on the structural property and thickness of $CuO_x$ and its interface with FMs. We postulate that different oxidation conditions can lead to different orbital textures that influence the sign of the torque. Moreover, the interplay between bulk and interface-based processes, namely the OHE and the orbital REE, can also vary based on differences in the interface quality (roughness, intermixing etc.) and whether or not the Cu layer is fully oxidized or a Cu-CuOx interface is still present. It is also worth noticing that $\xi_{DL}^{E}$ of the reference Py/Pt is about twice that of Co/Pt. This cannot be attributed to a difference in the quality of the Pt growth on different FMs, because the same magnitude of torque is obtained when the stacking sequence of the layers is reversed (square dots in Figure 2a-b and Figure 3a-b). Therefore, this indicates an intrinsic difference in the spin transparency of the Py/NM and Co/NM interfaces, favoring the former for both spin and orbital torque applications. The critical role of the spin transparency at the FM/NM interface is also evident when utilizing Ni as FM layer. Here, we observe comparable torques between Ni/$CuO_x$ and Ni/Pt, as expected of the large orbital-to-spin conversion efficiency of Ni. However, the overall (S)OTs efficiency is about two orders of magnitude lower with respect to the Py-based systems (see Supporting Information Note 7) due presumably to the low spin/orbital transmission at the Ni/NM interface.

The sign change of $\xi^E_{DL}$ between Co/Cu and Co/$CuO_x$ bilayers, arising from turning on and off the OHE in Cu by creating an orbital texture when the metal is oxidized, can be exploited to tune the OT via full-electric control of the oxidation state of Cu. To demonstrate this, we fabricated solid-state gated devices, where voltage-driven $O^{2-}$ ion migration across a $GdO_x$ insulating barrier can controllably and reversibly oxidize Cu. This approach is readily used in magnetoionics to modulate magnetic[38] and interface properties of thin films[39,40] and it has been applied to engineering of SOTs via oxygen manipulation.[41,42] However, electrical control of orbital currents has so far only been shown indirectly via measurement of the spin Hall magnetoresistance and through the use of liquid ion gating,[20] which is incompatible with implementation in microelectronics. Therefore, our experiments offer a more direct and scalable approach to technological applications.

Figure 4a shows a micrograph of a representative device overlayed with simplified electrical connections. We used $GdO_x$(25 nm) as the dielectric barrier and Pt(5 nm) as the top gate (Figure 4b). We note that Co(5)/$CuO_x$(3) was prepared using the same protocol as the devices measured earlier, hence Cu is naturally oxidized. When a positive gate voltage ($V_G$) is applied across the $GdO_x$, we expect negatively charged $O^{2-}$ ions to be attracted towards the top Pt contact (Figure 4b),[43] deoxidizing $CuO_x$. A negative biasing will then drive the $O^{2-}$ ions from $GdO_x$ into the Cu layer and oxidize the Cu.

Figure 4c displays the current vs. voltage ($I - V$) characteristic of a prototypical device under gating. As expected for an insulating barrier, the current is nearly zero at low gate voltages and becomes highly non-linear above $V_G \approx 3.5$V, indicating non-ohmic resistance and no apparent leakage across the 25 nm thick $GdO_x$. Based on this curve, we choose $V_G = \pm 3$V as the standard value for our experiments.

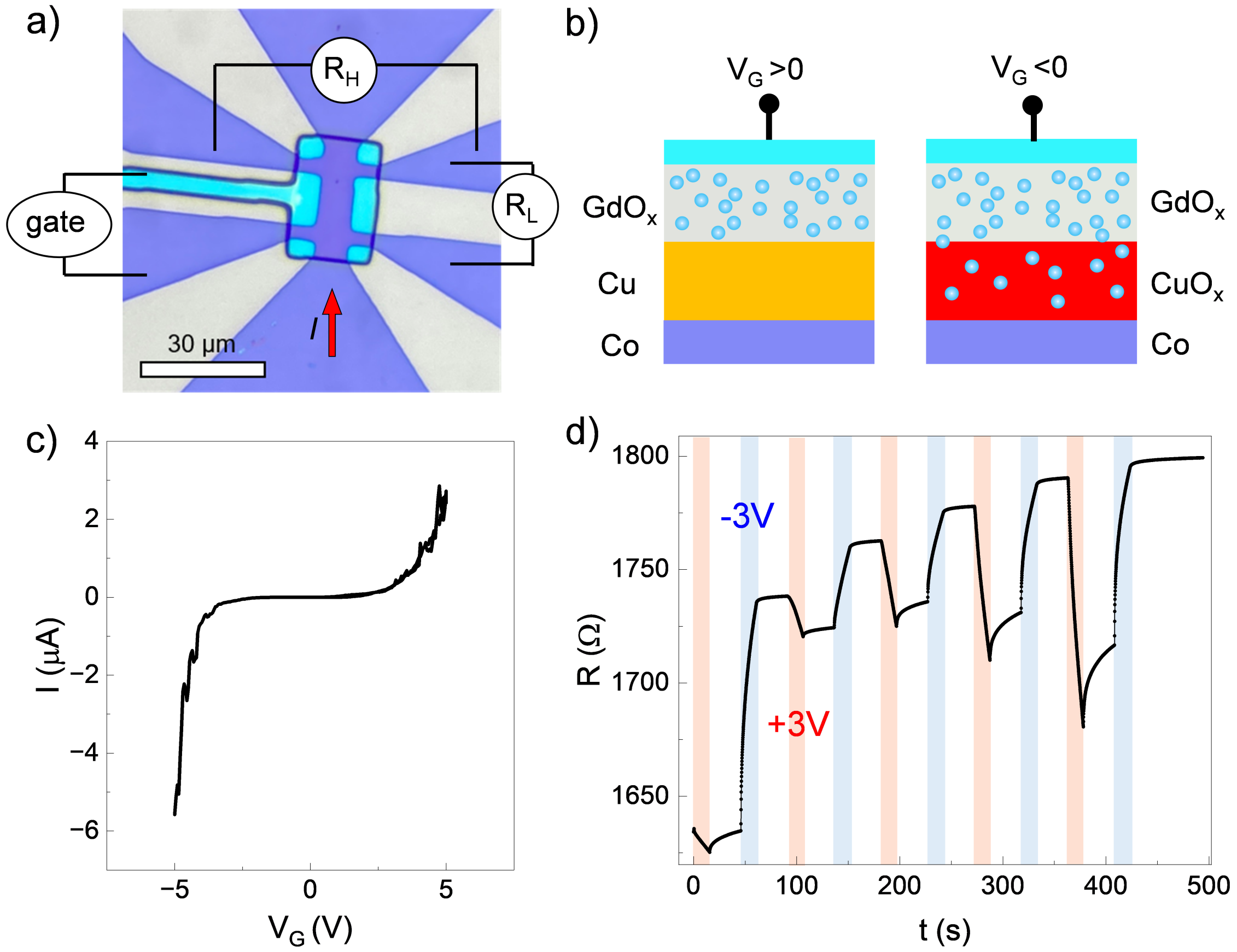

Figure 4: Solid-state gated devices. a) Optical microscope image of the device and schematics of the connections. b) Schematic representation of the voltage-drive $O^{2-}$ ions (blue spheres) migration across the cross-section of the sample. c) Plot of current ($I$) as a function of gate voltage measured in 2-point geometry between -5 and 5 V. d) Plot of longitudinal resistance $R_L$ as a function of time during application of a positive (red), negative (blue) or zero (white) gate voltage.

The voltage-driven $O^{2-}$ ions migration is reflected in a change of the resistance of the stack as (part of the) Cu looses or regains its metallic character. Figure 4d displays the evolution of the resistance ($R$) measured in the 2-point geometry of a prototypical device when gated. Positive bias (red) decreases $R$, while negative bias (blue) increases it. The relative change in $R$ increases after each gate iteration, suggesting a training effect owing to the ion mobility increasing after several gating cycles. Moreover, when no $V_G$ is applied (white), $R$ tends to increase slightly and then saturate, especially after positive biasing. This can be explained by an initial partial reduction of Gd and its natural tendency to reoxidize once the gate voltage is removed. On the other hand, once Gd gets reoxidized, $R$ remains essentially constant over the experimental time frame (see Supporting Information Note 8), indicating that voltage-driven ion migration is an effective means to the stable reduction of $CuO_x$ to metallic Cu.

Figure 5 shows the OT efficiency (defined in eq.4) measured at the three stages of the gating cycle for two different devices. In these experiments, $V_G$ is applied following the sequence 0V → +3V → 0V → -3V → 0V. In both cases, $\xi^E_{DL}$ is negative in the pristine state, confirming that the $CuO_x$ layer is initially oxidized (details on the fabrication process in Supplementary Material Note 1). In Device 1. a positive voltage of 3 V is applied for an arbitrarily long time of 750s, during which $R$ decreases $\approx$ 3.5%. After the positive gate, the torque sign becomes positive, as expected when the spin accumulation originates in metallic Cu. It is worth mentioning that the upon negative gating a $Cu/GdO_x$ is created which could exhibit the orbital REE, thereby producing an additional positive component to the torque.[26] Subsequently, we apply a negative $V_G$ for 1250s, after which $R$ increases $\approx$ 5% compared to its initial value in the pristine state. Measuring the OTs after negative gating reveals that the sign of the damping-like torque is reversed back to negative due to the oxidation of Cu.

Varying the duration of $V_G$ allows us to tune the oxidation state of Cu continuously and, hence, the magnitude of the damping-like torque as exemplified in Device 2. Applying a positive gate voltage of 1000s followed by a negative one of 2000s leads to almost complete suppression of the net torque. The $R_H$ vs $H_Z$ loops at the various stages of the experiment have been also measured. Here we observe that the amplitude of the AHE changes, owning to the change in current density across the FM, but $B_S$ remains essentially unvaried. Because $B_S$ is directly proportional to $M_S$, it can be used as an indication of potential oxidation of the FM layer. In the case of our experiment me conclude that a small interfacial oxidation of Co, if any, doesn't affect the amplitude or sign of the torque (see Supplementary Information Note 8).[44,45]

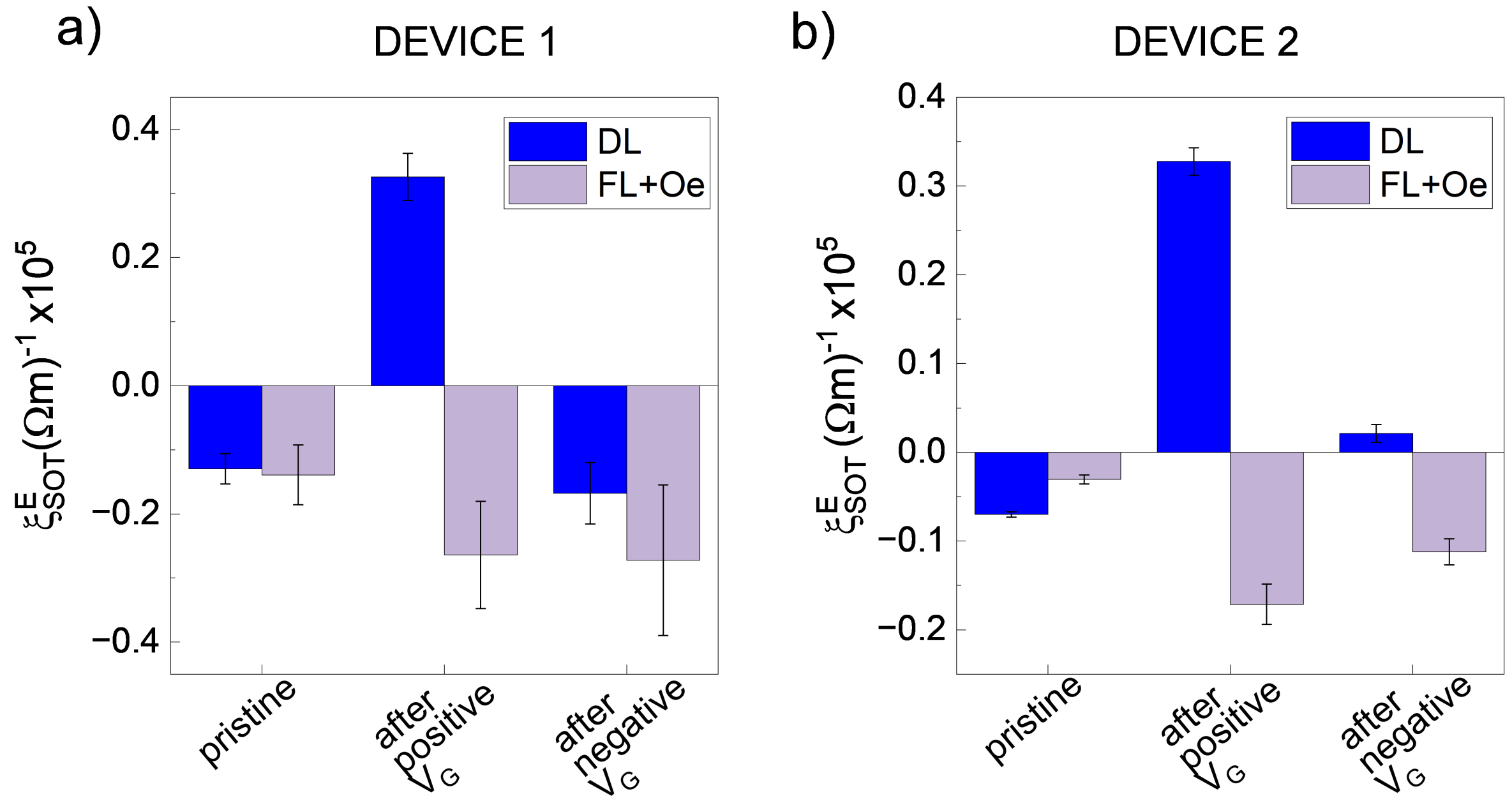

Figure 5: Electrical control of orbital torques. Plot of the damping-like ($\xi^E_{DL}$) (blue) and field-like + Oersted ($\xi^E_{FL+Oe}$) (violet) torque efficiency measured before gating (pristine) and after application of a positive and negative gate voltage ($V_G$). Gate cycle a) with and b) without sign reversal of the damping-like torque efficiency.

The field-like OT, differently from the damping-like component, remains negative across the gating sequence, as expected in the case of 5 nm thick Co (see Figure 2c). However, its magnitude increases as the resistance decreases, owing to the larger current density in the Cu layer and, thus, larger Oersted contribution. This further confirms the different origins of $B_{FL}$ and $B_{DL}$ in Co/Cu and Co/$CuO_x$ structures.

We have repeated similar gating experiments in multiple devices over multiple samples and verified that the general behavior is reproducible. However, we also observed large variability in the exact valued of $\xi^E_{DL}$. We attribute these differences to the inhomogeneity of the natural oxidation of Cu that is also reflected in differences in the devices resistance and gating times required for the oxidation/reduction of Cu. Further optimization is needed before this concept can be implemented for further application. Overall, our experiments demonstrate the feasibility of voltage-driven ion migration to control the oxidation state of Cu. Despite long gating times limited by the slow dynamics of the $O^{2-}$ ions, Figure 5 is a clear proof of concept for the possibility of dynamical control of OTs in spin-orbitronic devices.

In conclusion, we have studied the (S)OTs in Co($t_{Co}$)/NM and Py($t_{Py}$)/NM bilayers (NM = Py, Cu, $CuO_x$) utilizing harmonic Hall effect measurements. Comparing the results between different NM and analyzing the FM thickness dependence of the (S)OTs, we conclude that the damping-like component predominantly originates from the SHE in Pt and the OHE in $CuO_x$. Owning to the large orbital-to-spin conversion coefficient in Py, the OT in Py/$CuO_x$ is comparable in magnitude to the reference Py/Pt bilayer and exceeds that found in Co/Pt. This confirms that the OHE in this material is large and can be harnessed in spin orbitronics devices based on light metals. Moreover, the damping-like torque efficiency changes sign based on the oxidation state of Cu. We exploit this to demonstrate reversible tuning of the OT in Co(5)/$CuO_x$(3) via voltage-driven ion migration in solid-state gated devices. Our study is a major step towards harnessing orbital currents in cost-effective light elements to achieve efficient and dynamic electrical control of magnetization in spin-orbitronics devices.

## Supporting Information Available

Supporting Information: Methods; XPS analysis; VSM measurements; Parallel resistor model of current density; Self-induced torques; Spin and orbital diffusion length; Spin transparency of the FM/NM interface; Voltage-driven ion migration in Cu.

## Acknowledgement

S.D. acknowledges funding from the Marie Sklodowska-Curie Actions (MSCA) under the European Union's Horizon Europe research and innovation programme (project SPINDY, Grant Agreement No. 101106885). C.O.A. acknowledges funding from the European Research Council (ERC) under the European Union's Horizon 2020 research and innovation programme (project MAGNEPIC, Grant Agreement No. 949052). We acknowledge financial support from the State Investigation Agency, through the Severo Ochoa Programme for Centres of Excellence in R&D (CEX2023-001263-S). We thank Loes Fransen for initial tests on gated structures and fruitful discussions.

## References

(1) Manchon, A.; Železný, J.; Miron, I. M.; Jungwirth, T.; Sinova, J.; Thiaville, A.; Garello, K.; Gambardella, P. Current-induced spin-orbit torques in ferromagnetic and antiferromagnetic systems. *Reviews of Modern Physics* **2019**, *91*.

(2) Dieny, B. et al. Opportunities and challenges for spintronics in the microelectronics industry. *Nature Electronics* **2020**, *3*, 446–459.

(3) Kim, J.; Otani, Y. Orbital angular momentum for spintronics. *Journal of Magnetism and Magnetic Materials* **2022**, *563*.

(4) Rappoport, T. G. First light on orbitronics as a viable form of electronics. *Nature* **2023**, *619*.

(5) Go, D.; Jo, D.; Lee, H. W.; Kläui, M.; Mokrousov, Y. Orbitronics: Orbital currents in solids. *Europhysics Letters* **2021**, *135*.

(6) Go, D.; Jo, D.; Kim, C.; Lee, H. W. Intrinsic Spin and Orbital Hall Effects from Orbital Texture. *Physical Review Letters* **2018**, *121*, 086602.

(7) Kontani, H.; Tanaka, T.; Hirashima, D. S.; Yamada, K.; Inoue, J. Giant orbital hall effect in transition metals: Origin of large spin and anomalous hall effects. *Physical Review Letters* **2009**, *102*.

(8) Jo, D.; Go, D.; Lee, H. W. Gigantic intrinsic orbital Hall effects in weakly spin-orbit coupled metals. *Physical Review B* **2018**, *98*.

(9) Lee, S.; Kang, M. G.; Go, D.; Kim, D.; Kang, J. H.; Lee, T.; Lee, G. H.; Kang, J.; Lee, N. J.; Mokrousov, Y.; Kim, S.; Kim, K. J.; Lee, K. J.; Park, B. G. Efficient conversion of orbital Hall current to spin current for spin-orbit torque switching. *Communications Physics* **2021**, *4*.

(10) Sala, G.; Gambardella, P. Giant orbital Hall effect and orbital-to-spin conversion in 3d, 5d, and 4f metallic heterostructures. *Physical Review Research* **2022**, *4*.

(11) Huang, Q. K.; Liu, S.; Yang, T.; Xie, R.; Cai, L.; Cao, Q.; Lü, W.; Bai, L.; Tian, Y.; Yan, S. Current-Induced Magnetization Switching in Light-Metal-Oxide/Ferromagnetic-Metal Bilayers via Orbital Rashba Effect. *Nano Letters* **2023**, *23*, 11323–11329.

(12) Kim, J. y.; Cramer, J.; Lee, K.; Han, D. S.; Go, D.; Salev, P.; Lapa, P. N.; Vargas, N. M.; Schuller, I. K.; Mokrousov, Y.; Jakob, G.; Kläui, M. Tuning Spin-Orbit Torques Across the Phase Transition in VO2/NiFe Heterostructure. *Advanced Functional Materials* **2022**, *32*.

(13) Choi, Y. G.; Jo, D.; Ko, K. H.; Go, D.; Kim, K. H.; Park, H. G.; Kim, C.; Min, B. C.; Choi, G. M.; Lee, H. W. Observation of the orbital Hall effect in a light metal Ti. *Nature* **2023**, *619*, 52–56.

(14) Li, T.; Liu, L.; Li, X.; Zhao, X.; An, H.; Ando, K. Giant Orbital-to-Spin Conversion for Efficient Current-Induced Magnetization Switching of Ferrimagnetic Insulator. *Nano Letters* **2023**, *23*, 7174–7179.

(15) Hayashi, H.; Ando, K. Orbital Hall magnetoresistance in Ni/Ti bilayers. *Applied Physics Letters* **2023**, *123*.

(16) An, H.; Kageyama, Y.; Kanno, Y.; Enishi, N.; Ando, K. Spin-torque generator engineered by natural oxidation of Cu. *Nature Communications* **2016**, *7*.

(17) Karube, S.; Kondou, K.; Otani, Y. C. Experimental observation of spin-to-charge current conversion at non-magnetic metal/Bi2O3 interfaces. *Applied Physics Express* **2016**, *9*.

(18) Ding, S.; Ross, A.; Go, D.; Baldrati, L.; Ren, Z.; Freimuth, F.; Becker, S.; Kammerbauer, F.; Yang, J.; Jakob, G.; Mokrousov, Y.; Kläui, M. Harnessing Orbital-to-Spin Conversion of Interfacial Orbital Currents for Efficient Spin-Orbit Torques. *Physical Review Letters* **2020**, *125*.

(19) Xiao, Z. Y.; Li, Y. J.; Zhang, W.; Han, Y. J.; Li, D.; Chen, Q.; Zeng, Z. M.; Quan, Z. Y.; Xu, X. H. Enhancement of torque efficiency and spin Hall angle driven collaboratively by orbital torque and spin-orbit torque. *Applied Physics Letters* **2022**, *121*.

(20) An, T.; Cui, B.; Zhang, M.; Liu, F.; Cheng, S.; Zhang, K.; Ren, X.; Liu, L.; Cheng, B.; Jiang, C.; Hu, J. Electrical Manipulation of Orbital Current Via Oxygen Migration in Ni81Fe19/CuOx/TaN Heterostructure. *Advanced Materials* **2023**, *35*.

(21) An, T.; Cui, B.; Liu, L.; Zhang, M.; Liu, F.; Liu, W.; Xie, J.; Ren, X.; Chu, R.; Cheng, B.; Jiang, C.; Hu, J. Enhanced Spin Current in Ni81Fe19/Cu–CuOx Bilayer with Top and Sideways Oxidization. *Advanced Materials* **2023**, *35*.

(22) Ding, S.; Wang, H.; Legrand, W.; Noël, P.; Gambardella, P. Mitigation of Gilbert Damping in the CoFe/CuOx Orbital Torque System. *Nano Letters* **2024**, *4*, 32.

(23) Zheng, K.; Cao, C.; Lu, Y.; Meng, J.; Pan, J.; Zhao, Z.; Xu, Y.; Shang, T.; Zhan, Q.

Enhanced torque efficiency in ferromagnetic multilayers by introducing naturally oxidized Cu. *Applied Physics Letters* **2024**, *124*.

(24) Chen, T. Y.; Hsiao, Y. C.; Liao, W. B.; Pai, C. F. Tailoring Neuromorphic Switching by Cu Nx -Mediated Orbital Currents. *Physical Review Applied* **2022**, *17*.

(25) Kim, J.; Go, D.; Tsai, H.; Jo, D.; Kondou, K.; Lee, H. W.; Otani, Y. Nontrivial torque generation by orbital angular momentum injection in ferromagnetic-metal/Cu/Al2O3 trilayers. *Physical Review B* **2021**, *103*.

(26) Kim, J.; Uzuhashi, J.; Horio, M.; Senoo, T.; Go, D.; Jo, D.; Sumi, T.; Wada, T.; Matsuda, I.; Ohkubo, T.; Mitani, S.; Lee, H. W.; Otani, Y. Oxide layer dependent orbital torque efficiency in ferromagnet/Cu/oxide heterostructures. *Physical Review Materials* **2023**, *7*.

(27) Ding, S.; Noël, P.; Krishnaswamy, G. K.; Gambardella, P. Unidirectional orbital magnetoresistance in light-metal-ferromagnet bilayers. *Physical Review Research* **2022**, *4*.

(28) Sala, G.; Wang, H.; Legrand, W.; Gambardella, P. Orbital Hanle Magnetoresistance in a 3d Transition Metal. *Physical Review Letters* **2023**, *131*.

(29) Krishnia, S.; Bony, B.; Rongione, E.; Vicente-Arche, L. M.; Denneulin, T.; Pezo, A.; Lu, Y.; Dunin-Borkowski, R. E.; Collin, S.; Fert, A.; George, J. M.; Reyren, N.; Cros, V.; Jaffrès, H. Quantifying the large contribution from orbital Rashba-Edelstein effect to the effective damping-like torque on magnetization. *APL Materials* **2024**, *12*.

(30) Lee, D. et al. Orbital torque in magnetic bilayers. *Nature Communications* **2021**, *12*.

(31) Avci, C. O.; Garello, K.; Nistor, C.; Godey, S.; Ballesteros, B.; Mugarza, A.; Barla, A.; Valvidares, M.; Pellegrin, E.; Ghosh, A.; Miron, I. M.; Boulle, O.; Auffret, S.;

Gaudin, G.; Gambardella, P. Fieldlike and antidamping spin-orbit torques in as-grown and annealed Ta/CoFeB/MgO layers. *Physical Review B - Condensed Matter and Materials Physics* **2014**, *89*.

(32) Avci, C. O.; Garello, K.; Gabureac, M.; Ghosh, A.; Fuhrer, A.; Alvarado, S. F.; Gambardella, P. Interplay of spin-orbit torque and thermoelectric effects in ferromagnet/normal-metal bilayers. *Physical Review B - Condensed Matter and Materials Physics* **2014**, *90*.

(33) Garello, K.; Miron, I. M.; Avci, C. O.; Freimuth, F.; Mokrousov, Y.; Blügel, S.; Auffret, S.; Boulle, O.; Gaudin, G.; Gambardella, P. Symmetry and magnitude of spin-orbit torques in ferromagnetic heterostructures. *Nature Nanotechnology* **2013**, *8*, 587–593.

(34) Emori, S.; Nan, T.; Belkessam, A. M.; Wang, X.; Matyushov, A. D.; Babroski, C. J.; Gao, Y.; Lin, H.; Sun, N. X. Interfacial spin-orbit torque without bulk spin-orbit coupling. *Physical Review B* **2016**, *93*.

(35) Hayashi, M.; Kim, J.; Yamanouchi, M.; Ohno, H. Quantitative characterization of the spin-orbit torque using harmonic Hall voltage measurements. *Physical Review B - Condensed Matter and Materials Physics* **2014**, *89*.

(36) Wang, W. et al. Anomalous spin–orbit torques in magnetic single-layer films. *Nature Nanotechnology* **2019**, *14*, 819–824.

(37) Ochoa, H.; Zarzuela, R.; Tserkovnyak, Y. Self-induced spin-orbit torques in metallic ferromagnets. *Journal of Magnetism and Magnetic Materials* **2021**, *538*.

(38) Lee, K. Y.; Jo, S.; Tan, A. J.; Huang, M.; Choi, D.; Park, J. H.; Ji, H. I.; Son, J. W.; Chang, J.; Beach, G. S.; Woo, S. Fast magneto-ionic switching of interface anisotropy using yttria-stabilized zirconia gate oxide. *Nano Letters* **2020**, *20*, 3435–3441.

(39) Bauer, U.; Yao, L.; Tan, A. J.; Agrawal, P.; Emori, S.; Tuller, H. L.; Van Dijken, S.; Beach, G. S. Magneto-ionic control of interfacial magnetism. *Nature Materials* **2015**, *14*, 174–181.

(40) Herrera Diez, L. et al. Nonvolatile Ionic Modification of the Dzyaloshinskii-Moriya Interaction. *Physical Review Applied* **2019**, *12*.

(41) Qiu, X.; Narayanapillai, K.; Wu, Y.; Deorani, P.; Yang, D. H.; Noh, W. S.; Park, J. H.; Lee, K. J.; Lee, H. W.; Yang, H. Spin–orbit-torque engineering via oxygen manipulation. *Nature Nanotechnology 2015 10:4* **2015**, *10*, 333–338.

(42) Mishra, R.; Mahfouzi, F.; Kumar, D.; Cai, K.; Chen, M.; Qiu, X.; Kioussis, N.; Yang, H. Electric-field control of spin accumulation direction for spin-orbit torques. *Nature Communications 2019 10:1* **2019**, *10*, 1–8.

(43) Mishra, R.; Kumar, D.; Yang, H. Oxygen-Migration-Based Spintronic Device Emulating a Biological Synapse. *Physical Review Applied* **2019**, *11*.

(44) Feng, J.; Grimaldi, E.; Avci, C. O.; Baumgartner, M.; Cossu, G.; Rossi, A.; Gambardella, P. Effects of Oxidation of Top and Bottom Interfaces on the Electric, Magnetic, and Spin-Orbit Torque Properties of Pt / Co / Al Ox Trilayers. *Physical Review Applied* **2020**, *13*.

(45) Wu, S.; Jin, T. L.; Tan, F. N.; Ang, C. C.; Poh, H. Y.; Lim, G. J.; Lew, W. S. Enhancement of spin-orbit torque in Pt/Co/HfOx heterostructures with voltage-controlled oxygen ion migration. *Applied Physics Letters* **2023**, *122*, 122403.

# TOC Graphic

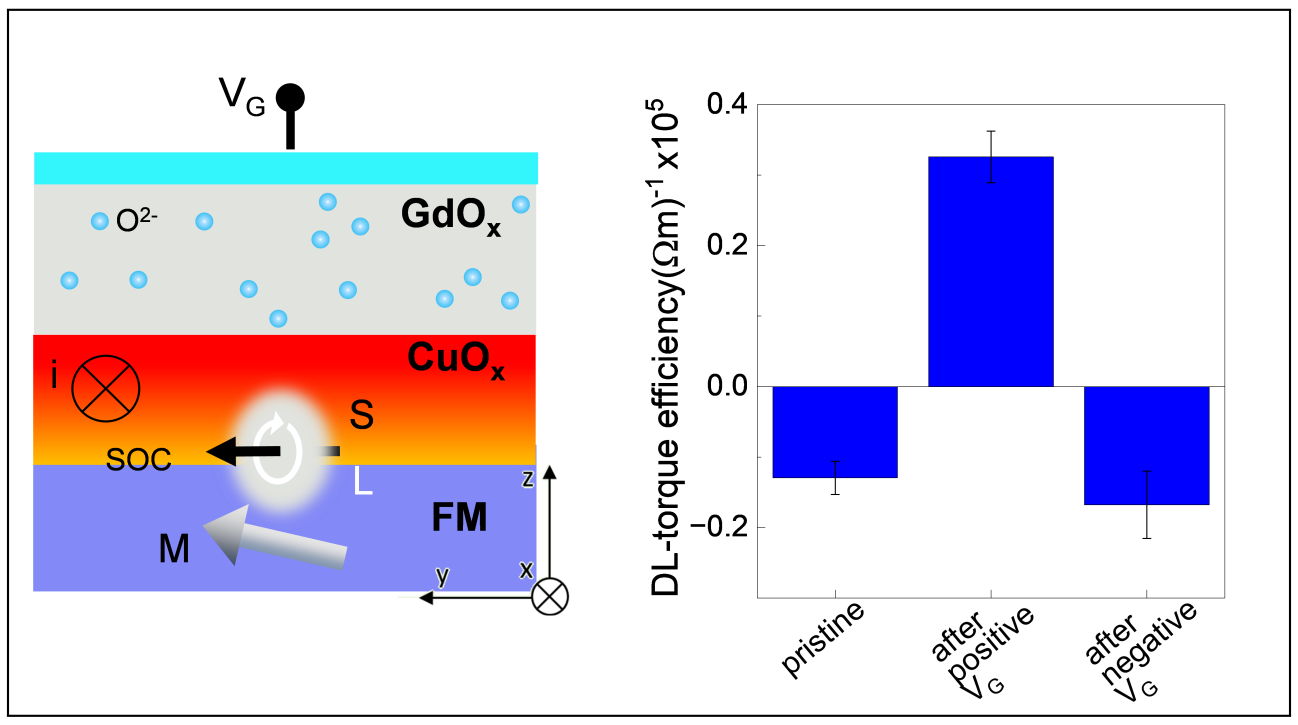

# Supporting Information for “Tunable spin and orbital torques in Cu-based magnetic heterostructures”

Silvia Damerio[1,*] and Can O. Avci[1]

[1]*Institut de Ciència de Materials de Barcelona, Campus de la UAB, Bellaterra, 08193, Spain*

*[*sdamerio@icmab.es](mailto:sdamerio@icmab.es)*

## Table of contents

## 1. Methods

**Sample Preparation.** Bilayers of Co/Cu, Co/Pt, Py/Cu and Py/Pt with in-plane magnetic anisotropy were deposited on thermally oxidized Si(001)/SiO2 substrates at room temperature by DC magnetron sputtering with 3 mTorr Ar. The deposition rates of Co, Py, Cu and Pt were 0.104 nms$^{-1}$ at 200 W , 0.0568 nms$^{-1}$ at 100 W, 0.0675 nm nms$^{-1}$ at 50 W and 0.185 nms$^{-1}$ 1 at 50 W, respectively. Adhesion and capping (where specified) Ti layers were also deposited at 200 W with a rate of 0.057 nms$^{-1}$. To obtain Co/CuO$_X$ and Py/CuO$_X$, the fabricated bilayers were left in air overnight to naturally oxidize Cu before the transport measurements. For the gated devices, after lift-off of the patterned metallic stack, a continuous layer of GdO$_X$ was deposited by RF magnetron sputtering with 3 mTorr Ar:O2 (30:5) at 50 W and deposition rate of 0.0064 nms$^{-1}$. Finally, Pt contacts and the top gate were patterned and deposited. The resistance through the GdO$_X$ spacer was > 100MΩ ensuring negligible crosstalk.

**Magnetic and Electrical Measurements.** The magnetic properties of the continuous layers were measured using vibrating sample magnetometry in a Quantum Design Physical Property Measurement System (PPMS). For transport measurements, the Co/CuO$_X$ and Py/CuO$_X$ films were patterned into six-terminal

Hall bar devices by standard laser-writer optical lithography and lift-off. The Hall bar dimensions are 30 μm for the current line length, 7.5 μm its width, and 3 μm the Hall branch width. The anomalous Hall voltage and angular-dependent first- and second-harmonic voltages were measured using a Zurich Instruments MFLI digital lock-in amplifier under an ac current $I = I_0 sin(\omega t)$ of amplitude $I_0$ = 6 − 12 mA and frequency ω = 2.9 kHz provided by a Keithley 6221 current source. For the gating experiments, a dc voltage of ±3V was applied between the top gate and one of the Hall arms with a Keithley 2470 Source Meter. All measurements were performed at room temperature.

**X-Ray diffraction**

The polycrystalline structure of Py films has been investigated by means of grazing incidence X-ray diffraction (GIXRD) in a Bruker D8-Discover diffractometer with Cu K-alpha radiation ($\lambda$ =1.5406 Å). In the spectrum, shown in Figure S1, we observed a single peak at $2\theta$=44.3º which agrees with the (111) diffraction peak of face-centered cubic (FCC) $Ni_{0.8}Fe_{0.2}$ with lattice parameter 3.55 Å [*A. V. Svalov et al., Sensors 22(21), 8357 (2022)*], indicating that the films are polycrystalline and do not deviate significantly from the nominal target stoichiometry (Fe 20% Ni 80%).

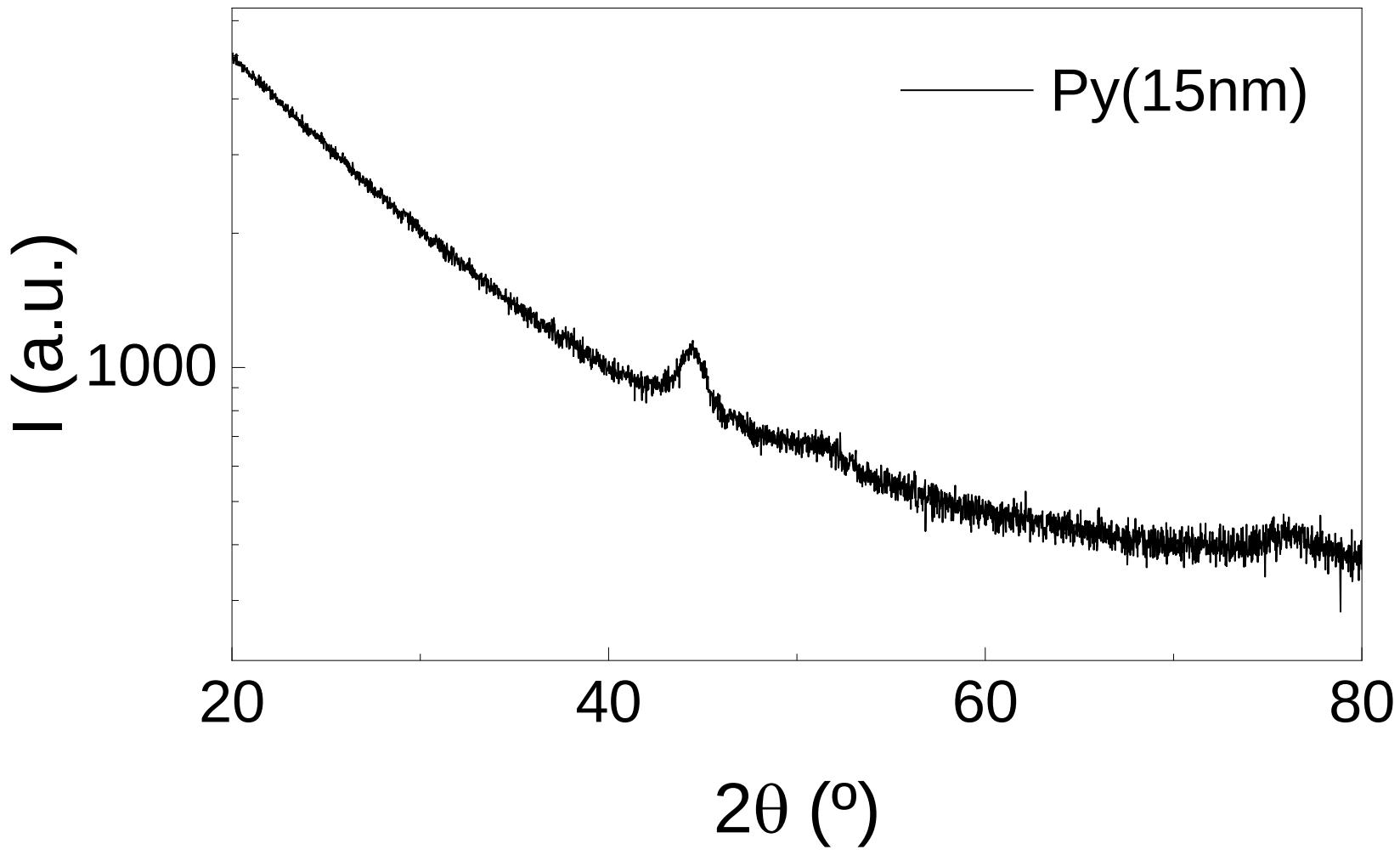

**Figure S1 Py crystal structure.** GIXRD pattern of a 15 nm thick Py film grown on Si substrate with a 1.5 nm Ti buffer layer.

## 2. XPS analysis

To characterize the composition of our materials, the oxidation state of Ti, O, Cu and Co was investigated by X-ray Photoelectron spectroscopy (XPS). We prepared three different samples: 1- Ti(1.5)/Co(5)/$CuO_X$ (3), where $CuO_X$ was obtained by natural oxidation of Cu, 2 - Ti(1.5)/Co(5)/Cu(3)/Ti(1.5) and 3- Ti(1.5)/$CoO_X$(6)/Cu/Ti(1.5), where $CuO_X$ was obtained by natural oxidation of Co. The spectra shown in Figure S2 were collected on a SPECS PHOIBOS 150 system and analyzed with the software CasaXPS. As it can be seen from the Ti 2p region (Figure S2a) the samples with Ti as top capping layer (red and blue) display a strong $TiO_2$ 2p peaks, while the noisier peaks in the oxidized specimen (gray) originate from the bottom Ti layer used for adhesion. The O 1s region of the Ti capped samples (Figure S2b) displays a single peak, indicating that all the Ti is uniformly oxidized. On the other hand, the presence of multiple peaks in the oxidized sample indicates that the oxidation of the top Cu layer is not uniform across the sample thickness. This is also reflected in the Cu region of the spectrum (Figure S2c), where the Cu 2p peaks of the oxidized sample is shifted towards higher binding energies compared to the other two samples. Finally, looking at the Co 2p region (Figure S2d), a significant difference is only found in peaks of the sample where Co was exposed to air, indicating that in the other two the oxidation doesn't reach Co.

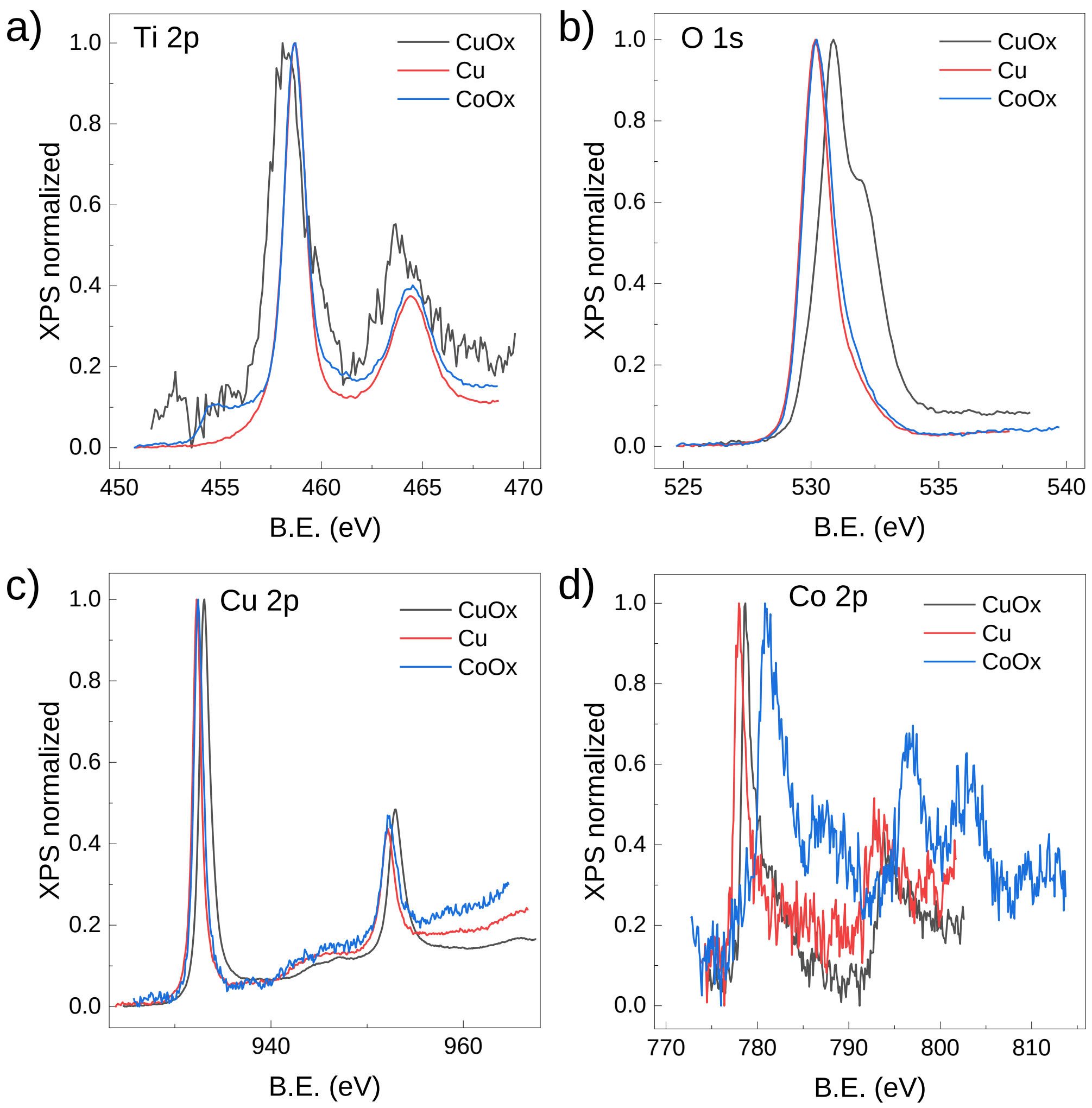

**Figure S2 XPS spectra.** a) Ti 2p, b) O 1s, c) Cu 2p and d) Co 2p regions of the XPS spectrum of a Ti(1.5)/Co(5)/$CuO_X$(3) (gray), Ti(1.5)/Co(5)/Cu(3)/Ti(1.5) (red) and a Ti(1.5)/$CoO_X$(6)/Cu(3)/Ti(1.5) (blue) samples.

## 3. VSM measurements

The magnetic properties of the FM(t)/Cu(3) bilayers were studied by means of vibrating sample magnetometry in a Quantum Design Physical Property Measurement System (PPMS). Figure S3a shows the in-plane magnetic hysteresis loops of Co samples with variable thickness. The saturation magnetization $M_S$ of Co ranges from $1.17\times10^6$ A/m for the 1.5 nm thick film to $1.25\times10^6$ A/m for 5 nm thick one. This reduction of $M_S$ compared to the bulk value of $1.44\times10^6$ A/m [*J.M.D. Coey, "Magnetism and Magnetic Materials" Cambridge University Press (2010)*], is not uncommon in thin films and is usually attributed to reduced dimensionality, microstructural variations and interface effects. Previous works *[C. O. Avci et al. Physical Review B 100, 235454 (2019); S. Lee et al.*, *Communications Physics 4, 234 (2021)]* report similar values for Co films below 10 nm and show that it also depends on the underlying layer.
Conversely, the $M_S$ of Py (Figure S3b) doesn't vary significantly with thickness and is $6.8\times10^5$ A/m for both 2.5 and 10 nm thick samples, close to bulk value.

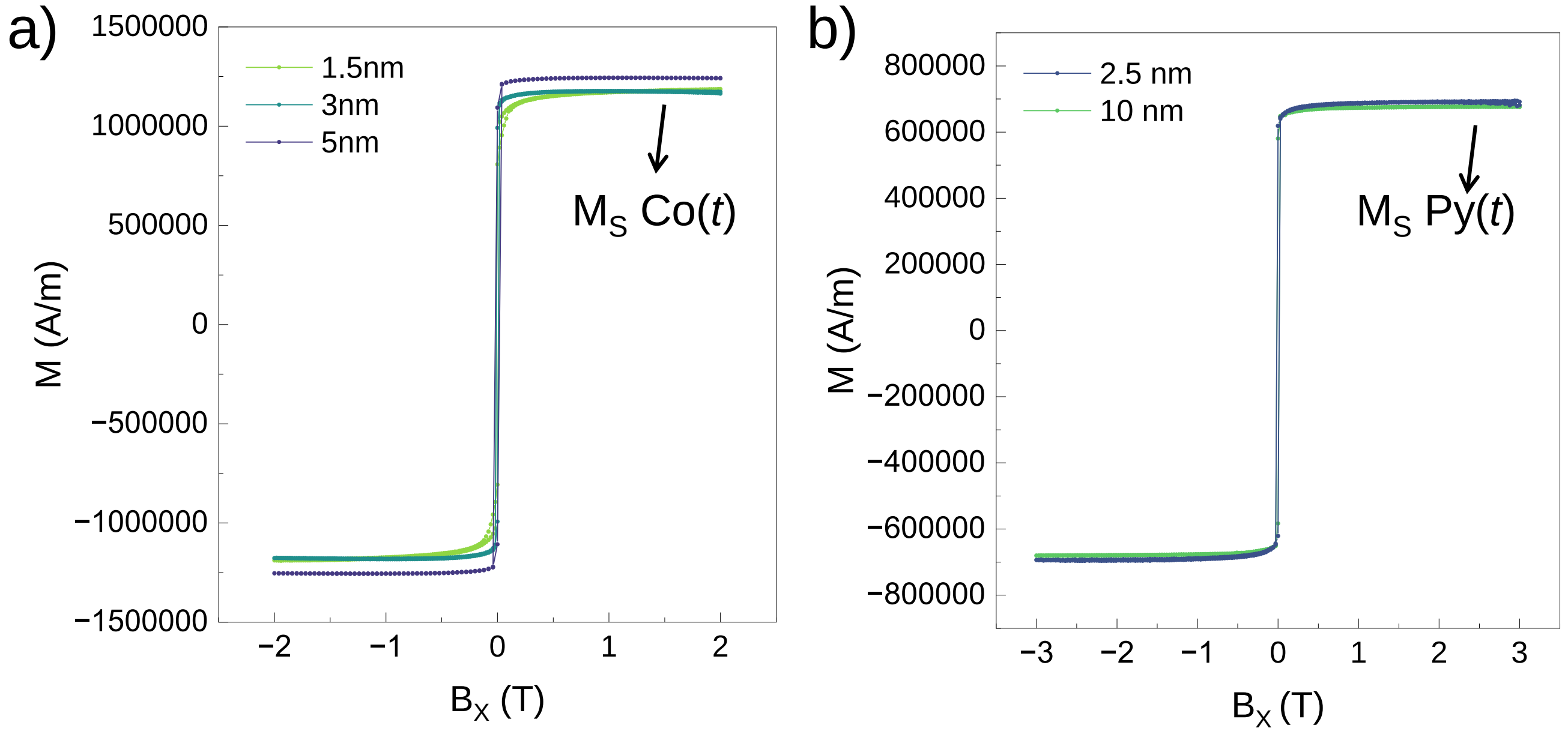

**Figure S3 M-H loops**. a) Plot of magnetization (*M*) as a function of in-plane field ($B_X$) of Co films with thickness of 1.5, 3 and 5 nm. b) Plot of *M* as a function of $B_X$ of Py films with thickness of 2.5 and 10 nm.

## 4. Parallel resistor model of current density

The resistivity of Co, Py, Cu and Pt are measured from reference single-layer samples and result to be 37.2 μΩ/cm, 51.8 μΩ/cm, 8.8 μΩ/cm and 50 μΩ/cm respectively. These values are comparable with what previously obtained for films of comparable thickness [*Y. Ke, et al., Phys. Rev. B 79, 155406 (2009), Nguyen M-*

*H. et al., Phys. Rev. Lett. 116, 126601 (2016)*]. Similar resistivity values for Co and Py can also be obtained from the fit of the FM layer thickness dependence of the inverse sheet resistance ($1/R_S$) of the samples, as shown in Figure S4.1a and Figure S4.1b. The resistivity of the individual layers is then used to calculate the current density in the NM, which is given by:

$$j_{NM} = \frac{i}{wt_{NM}} \frac{t_{NM}\rho_{FM}}{t_{NM}\rho_{FM} + t_{FM}\rho_{NM}}$$

where $i$ is the total current injected in the device and $w$=7.5 μm is the width of the Hall bar.

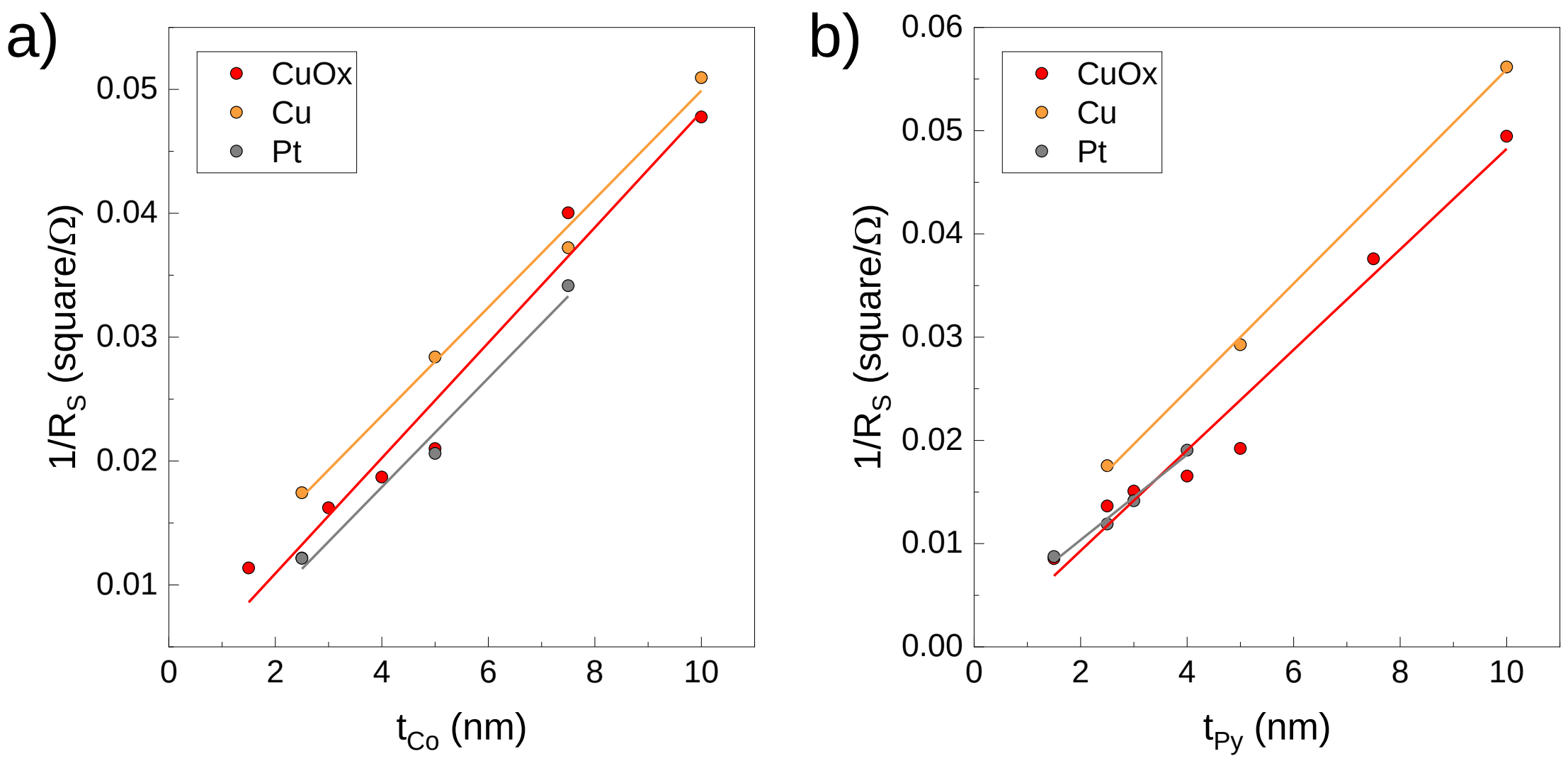

**Figure S4.1 Resistance of Co and Py.** Plot of the Inverse sheet resistance ($1/R_S$) as a function of a) Co thickness ($t_{Co}$) and b) Py thickness ($t_{Py}$) in Co(t)/NM(3) and Py(t)/NM(3) bilayers with NM=$CuO_x$ (red), NM=Cu (orange) and NM=Pt (gray).

The Oersted filed generated by the current flowing in the NM can be then calculated as using:

$$B_{Oe} = -\frac{\mu_0 i_{NM}}{2w}$$

where $\mu_0$ is the vacuum permeability and w the width of the Hall bar device. We used the values of $B_{Oe}$ obtained in this way to estimate $B_{FL}$ in Figures 2c) and 3c) of the main text.

Figure S4.2 shows the plot of the effective field of the field-like and Oersted component of the torque measured experimentally (open symbols) and the calculated $B_{Oe}$ (green lines) as a function of thickness of the FM. As it can be seen, the amplitude of $B_{Oe}$ decreases as expected with increasing $t_{FM}$, because the current in the NM decreases. The same trend is observed for $B_{FL+Oe}$ for FM/Pt bilayers. However, when $CuO_X$ is used as torque generator the trend reverses. This indicates the presence of a field-like component of the torque with opposite sign with respect to that generated in Pt that competes with the Oersted component.

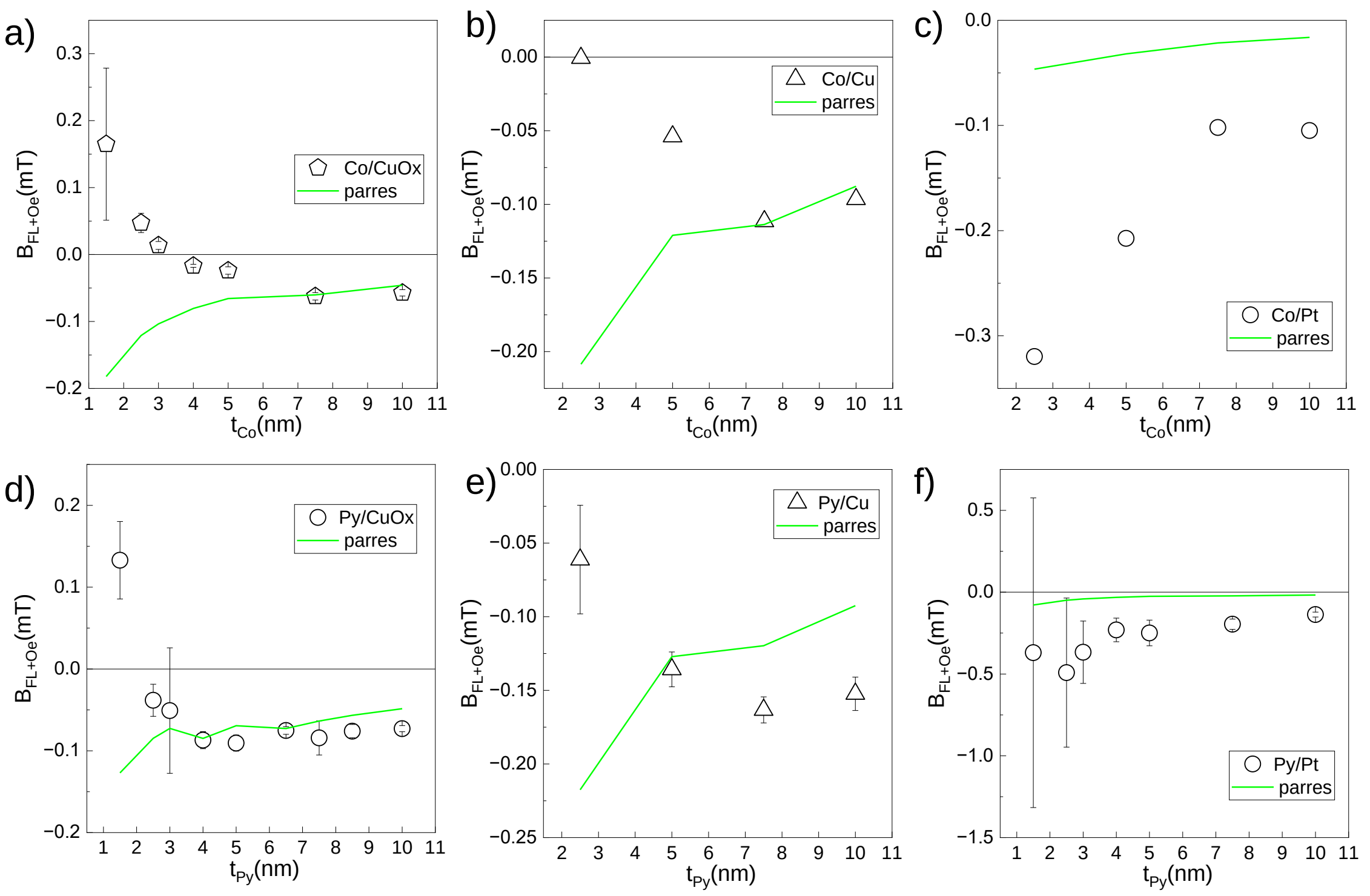

**Figure S4.2 Current induced Oersted field.** Plot of the field-like and Oersted effective field ($B_{FL+Oe}$) as a function of Co thickess ($t_{Co}$) for a) Co(t)/CuOx(3), b) Co(t)/Cu(3) and c) Co(t)/Pt(3) bilayers. Plot of $B_{FL+Oe}$ as a function of Py thickess ($t_{Py}$) for d) Py(t)/CuOx(3), e) Py(t)/Cu(3) and f) Py(t)/Pt(3) bilayers. The green lines represent the calculated $B_{Oe}$.

## 5. Self-induced torques

Spin accumulations can also be generated by spin-polarized currents in metallic ferromagnets without the active intervention of adjacent normal metals [*H. Ochoa, R. Zarzuela, and Y. Tserkovnyak, J. Magn. Magn. Mater. 538, 168262 (2021)*]. To estimate the magnitude of the self-induced torque in Co and Py we measured single-layer films. Figure S5a-b show the comparison of the torque on a 5 nm Co film with different capping layers. As apparent, both damping-like and field-like component of the torque are close to zero in the absence of an adjacent NM (self-torque). Small positive damping-like torque is observed for Cu capping layer, whereas $CuO_x$ provides a negative torque. A similar negative torque is observed when the Co sample is exposed to air before the deposition of the Cu(3)/Ti(1.5). This indicates that the formation of $CoO_x$ at the interface enhances the SOT efficiency, as observed in Co/$CoO_x$/Pt [*X. Feng, et al., Appl. Phys. Lett. 118, 132410 (2021)*]. The same is found for Py based bilayers (Figure S5c-d): no self-torque is observed without capping layer, but, differently from the Co case, we also found no significant enhancement of the SOT upon oxidation of Py.

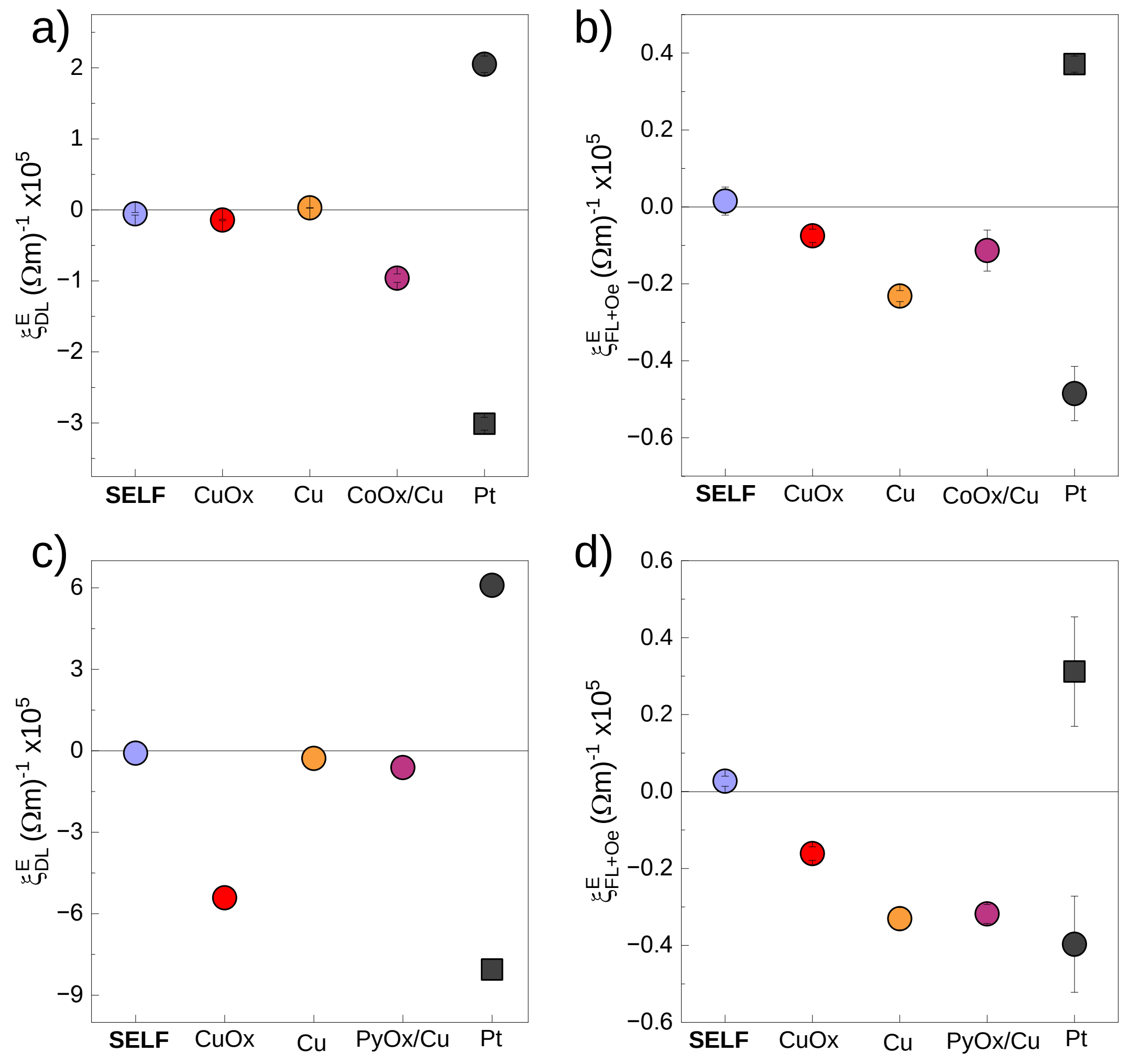

**Figure S5 Comparison of capping layer on SOTs**. a) Damping-like torque efficiency ($\xi^E_{DL}$) and b) field-like and Oersted torque efficiency ($\xi^E_{FL+Oe}$) in different Co(5)/NM bilayers. c) $\xi^E_{DL}$ and d) $\xi^E_{FL+Oe}$ in different Py(5)/NM bilayers.

## 6. Spin and orbital diffusion length

To quantify the spin diffusion length in Pt ($\lambda_{Pt}$) we studied the SOT in Co(5)/Pt($t$) bilayers. Figure S6.1a shows a plot of the damping-like component of the torque as a function of Pt thickness ($t_{Pt}$), which increases steeply for the first few nm and saturates above 3 nm. From the fit of the experimental points with an hyperbolic tangent function (green line) we estimate $\lambda_{Pt}$ in our system to be 1.56 nm, in good agreement with previous reports [*S. Ding et al., Nano Letters 24 (33), 10251–10257*

*(2024)*]. Figure S6.1b also shows the field-like and Oersted component of the torque as a function of Pt thickness ($t_{Pt}$). Here we see that the experimental points lie above the estimated Oersted component (black), thus indicating a positive field-like torque.

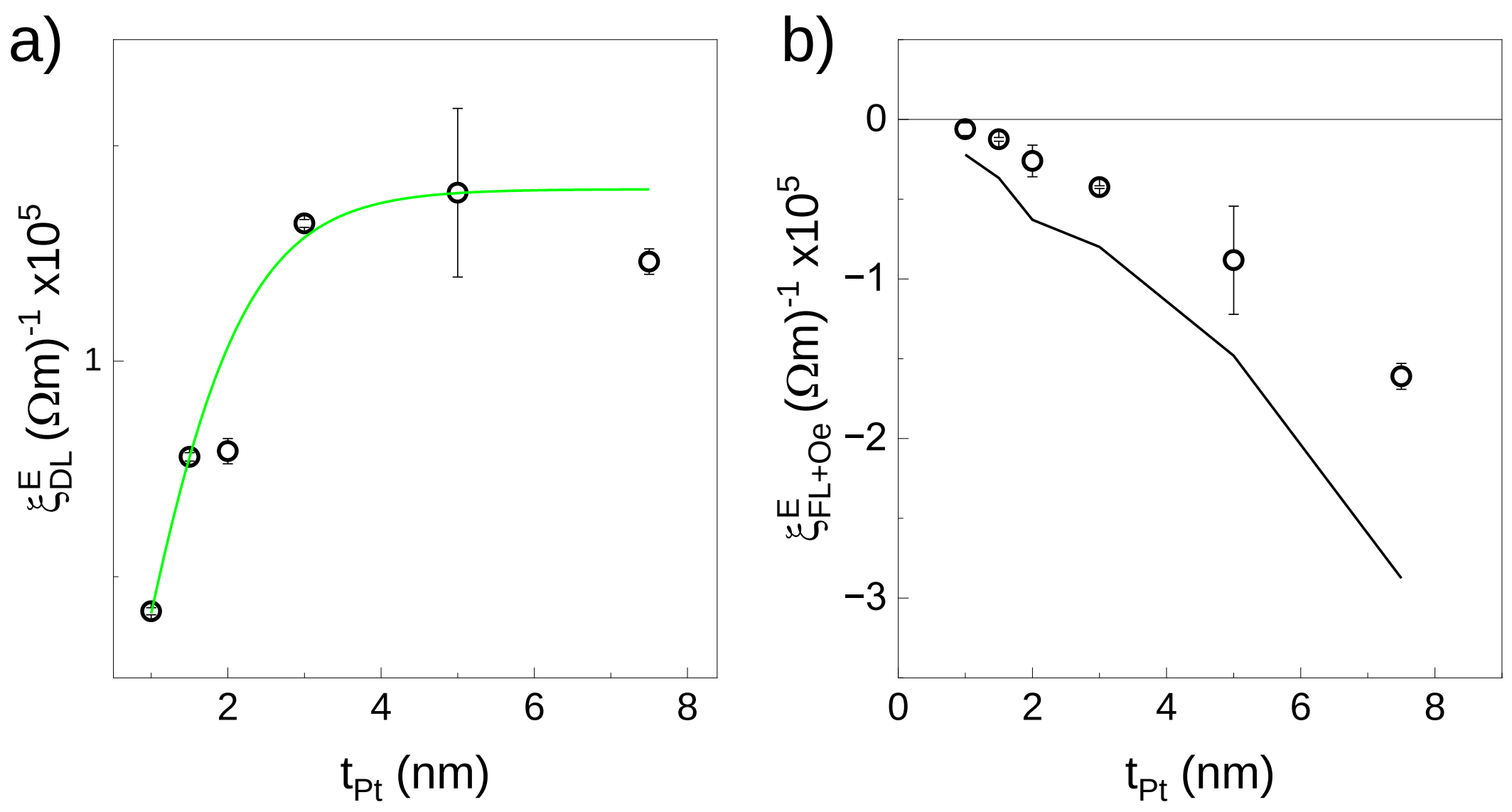

**Figure S6.1 Spin diffusion length in Co(5)/Pt(t) bilayers**. a) Damping-like torque efficiency ($\xi^E_{DL}$) and b) field-like and Oersted torque efficiency ($\xi^E_{FL+Oe}$) as a function of Pt thickness ($t_{Pt}$). The green solid line indicates the fit with a hyperbolic tangent function, while the black solid line indicates the estimated Oersted field generated by the current flow.

We have also made an attempt to estimate the saturation value of the orbital diffusion length in FM/CuOx(3) bilayers. To do so, we have fitted the data in Figure 3b of the manuscript using an hyperbolic tangent function (green in Figure S6.2a) of the type: *y=A*tanh(x/B+C)+D,* where $\lambda_{Py}$=*B+C*. Setting C=1.5 nm, the smallest value of $t_{Py}$ that we measured we obtain $\lambda_{Py}$≈4 nm. This value is somewhat lower than what found by Hayashi and coworkers *[H. Hayashi et al. Communication Physics 6,32 (2023)*] who have seen that the damping-like torque efficiency increases with $t_{FM}$ up to $t_{Ni}$ > 20 nm in Ni/Ti bilayers, but still larger than the spin diffusion length in Pt. We have also tentatively repeated the same fit for the data in Figure 2b of the manuscript and extract $\lambda_{Co}$, however the results of the fit (shown in Figure S6.2b) are in this case less reliable as we don't see any saturation of $\xi_{DL}$ up to the largest thickness. Therefore, in the case of Co we can only conclude that $\lambda_{Co}$≈4.5 nm only represents a lower limit to the actual orbital diffusion length

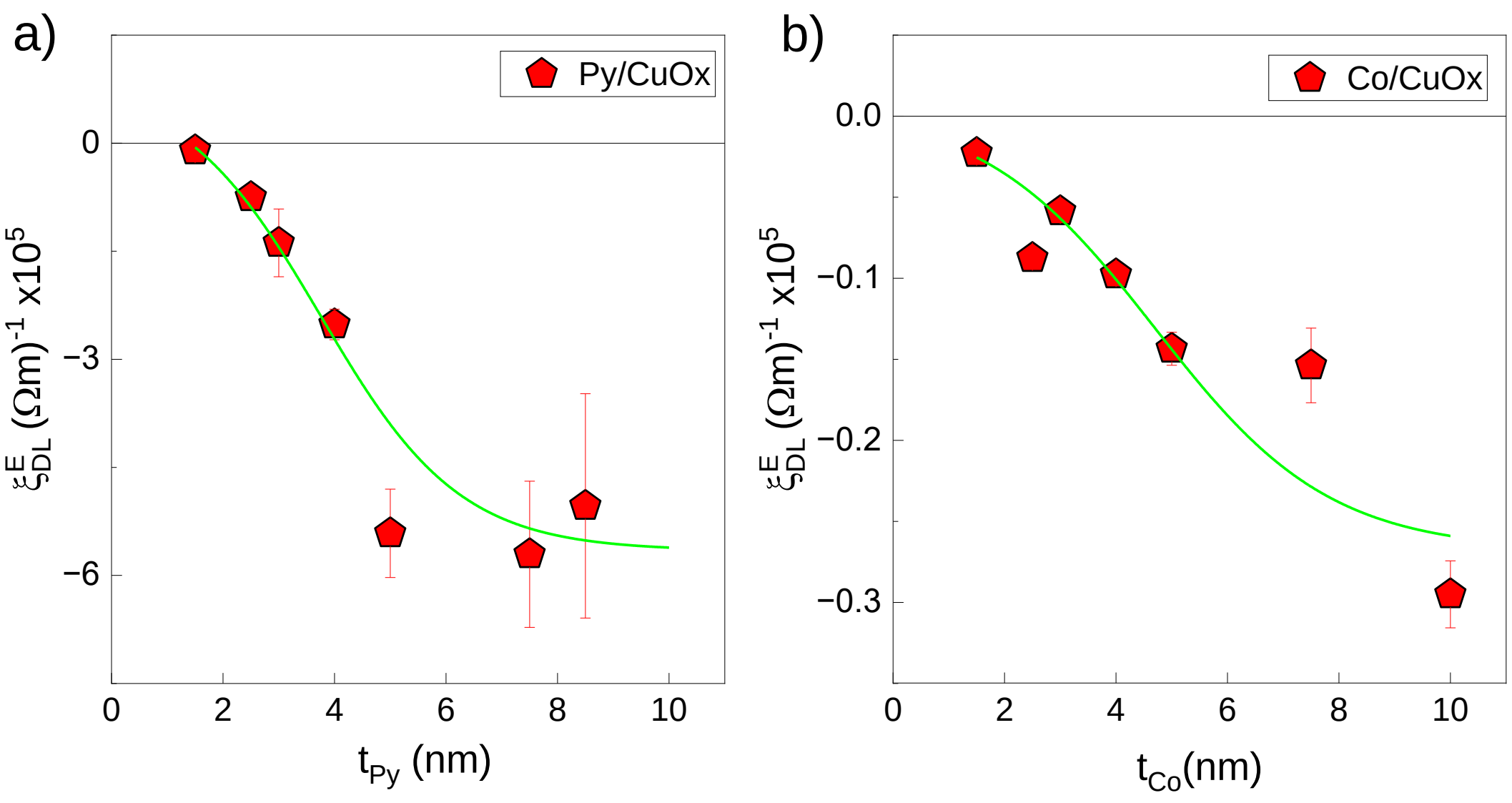

**Figure S6.2 Orbital diffusion length in FM/CuOx(3) bilayers**. a) Damping-like torque efficiency ($\xi^E_{DL}$) of a) Py thickness ($t_{Py}$). and b) Co thickness ($t_{Co}$). The green solid line indicates the fit with a hyperbolic tangent function.

## 7. Spin transparency at the FM/NM interface

It is well-known that the bulk of Ni is more efficient than Py in the orbital-to-spin conversion. Therefore, we attempted to measure the orbital torques in Ni/CuOx bilayers. Figure S7a shows the damping-like torque efficiency ($\xi_{DL}$) measured in different FM/NM bilayers. Surprisingly, the absolute value of $\xi_{DL}$ for Ni(5)/CuOx(3) (blue) is two orders of magnitude lower than in the Py/CuOx(3) case (green). However, because the same amplitude difference is found when comparing the two reference samples Ni(5)/Pt(3) and Py(5)/Pt(3), we conclude that the origin of the lower signal is related to the interfacial spin/orbital injection efficiency at the Ni/NM interfaces presumably due to more diffuse interface, making this material less suitable for harmonic Hall measurements. On the other hand, using Co as a FM, $\xi_{DL}$ with Pt is one order of magnitude larger than with $CuO_x$ (red in Figure S7a). The collective data corroborate our thesis that that Py and Ni are more efficient than Co for converting orbital currents that originate from $CuO_X$. For a better comparison of the three FM metals, Figure S7b shows a plot of the ratio between $\xi_{DL}$ measured in each FM/$CuO_X$ and corresponding FM/Pt reference. In view of the much larger amplitude of the signal, choosing Py over Ni is more sensible for the study of the FM thickness dependence of $\xi_{DL}$. Furthermore, we also noticed that the sign of $\xi_{DL}$ for Ni(5)/$CuO_X$(3) is positive. We hypothesize that this is related to the sign of the orbital-to-spin conversion in Ni, that could be opposite than that of Co and Py, in the same way as the AHE is also opposite in sign in Ni. Because both quantities depend on the interplay of their electronic band structure, Berry curvature, and spin-orbit coupling they could be related.

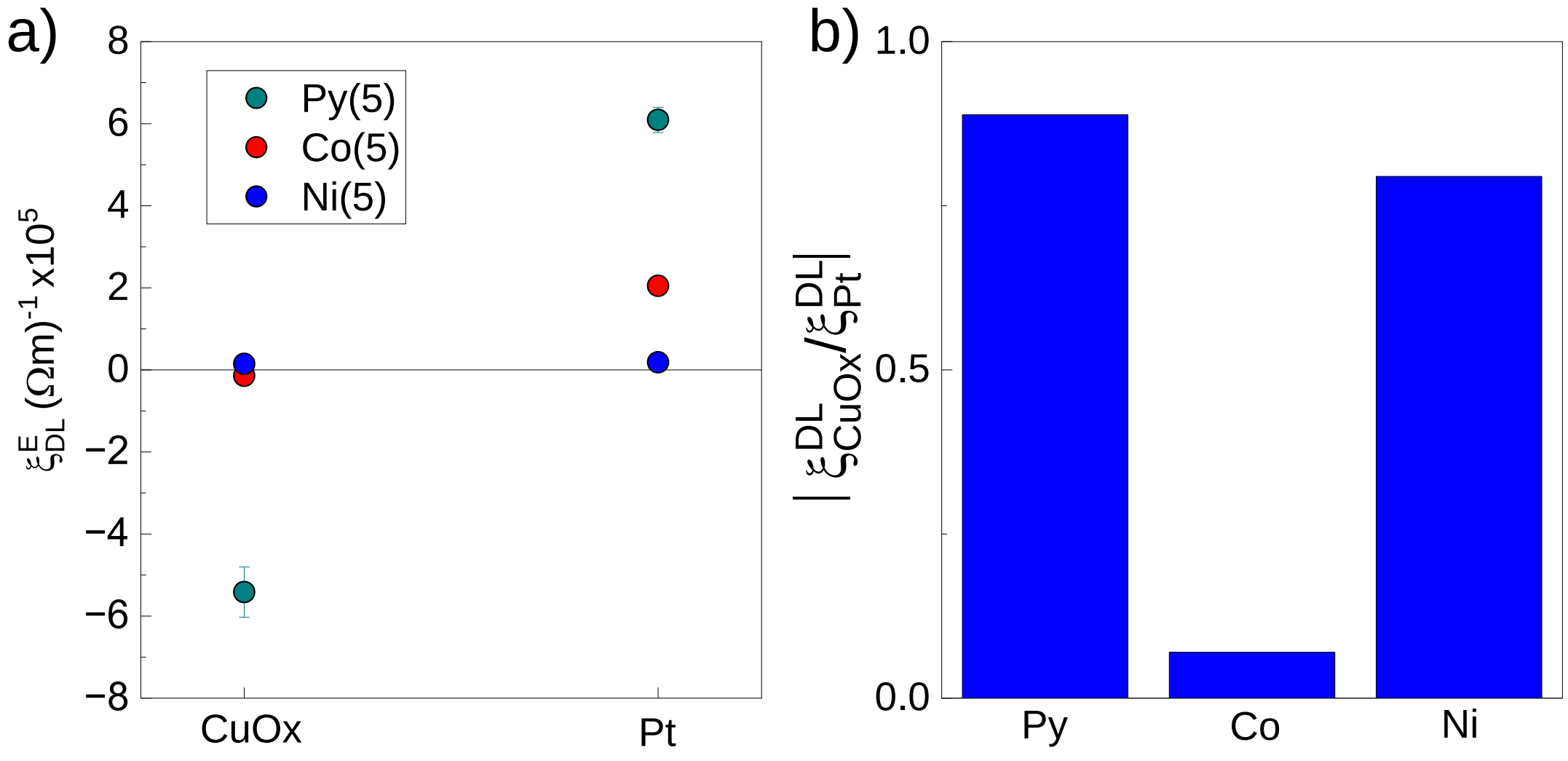

**Figure S7 Spin transparency of the FM/NM interface.** a) Plot of the damping-like torque efficiency ($\xi_{DL}$) in FM/$CuO_X$ and FM/Pt bilayers with FM=Py (green), FM=Co (red) and FM=Ni (blue). b) Plot of the absolute value of the ratio between $\xi_{DL}$ measured in each FM/$CuO_X$ and corresponding FM/Pt reference.

## 8. Voltage-driven ion migration in Cu

An indirect proof of the formation of metallic Cu via ion-migration upon gating with a positive gate voltage ($V_G$) comes from following the resistance ($R_L$) of the stack during the gating experiment. Upon applying a negative $V_G$, the Cu present in the stack gets oxidized and the resistance increases (light blue in Figure S8.1a). When $V_G$ is set to zero after negative bias $R_L$ remains stable, indicating that no further oxidation occurs. This is logical, as $CuO_X$ and $GdO_X$ are chemically stable at room temperature. On the other hand, upon applying a positive $V_G$ (light red in Figure S8.1a), the $O^{2-}$ ions migrate towards the top contact and out of $CuO_X$. Here the reduction process is not only limited to $CuO_X$ but part of $GdO_X$ can also loose oxygen. In this case, when gating is stopped after application of a positive $V_G$, $R_L$ shows an initial sudden upturn (gray arrow in Figure S8.1a). This is due to the spontaneous re-oxidation of Gd that has a large affinity to oxygen. However, after this initial steep increase $R_L$, it tends to stabilize at a value of resistance close to the minimum reached after gating. This indicates that the re-oxidation of Cu to $CuO_X$ is a much slower process. A similar behavior is also displayed by the device in Figure 4d of the manuscript. To further test the stability of the formation of metallic Cu we have monitored the evolution of the sample resistance with time for over 15 h after gating. As it can be seen in Figure S8.1b, after the initial increase due to the oxidation of Gd, $R_L$ increases slowly and never reaches its initial value. Therefore, the reduction of $CuO_X$ to Cu is stable within the time frame of out OT measurements (which only takes minutes) and therefore, we believe that voltage-driven ion migration is effective for comparing the OT generated in metallic Cu and $CuO_X$. Indeed, we cannot estimate which portion of the thickness of the $CuO_X$ layer gets reduced to metallic Cu, but the sign change of the torque indicates which species (Cu o $CuO_X$) predominates.

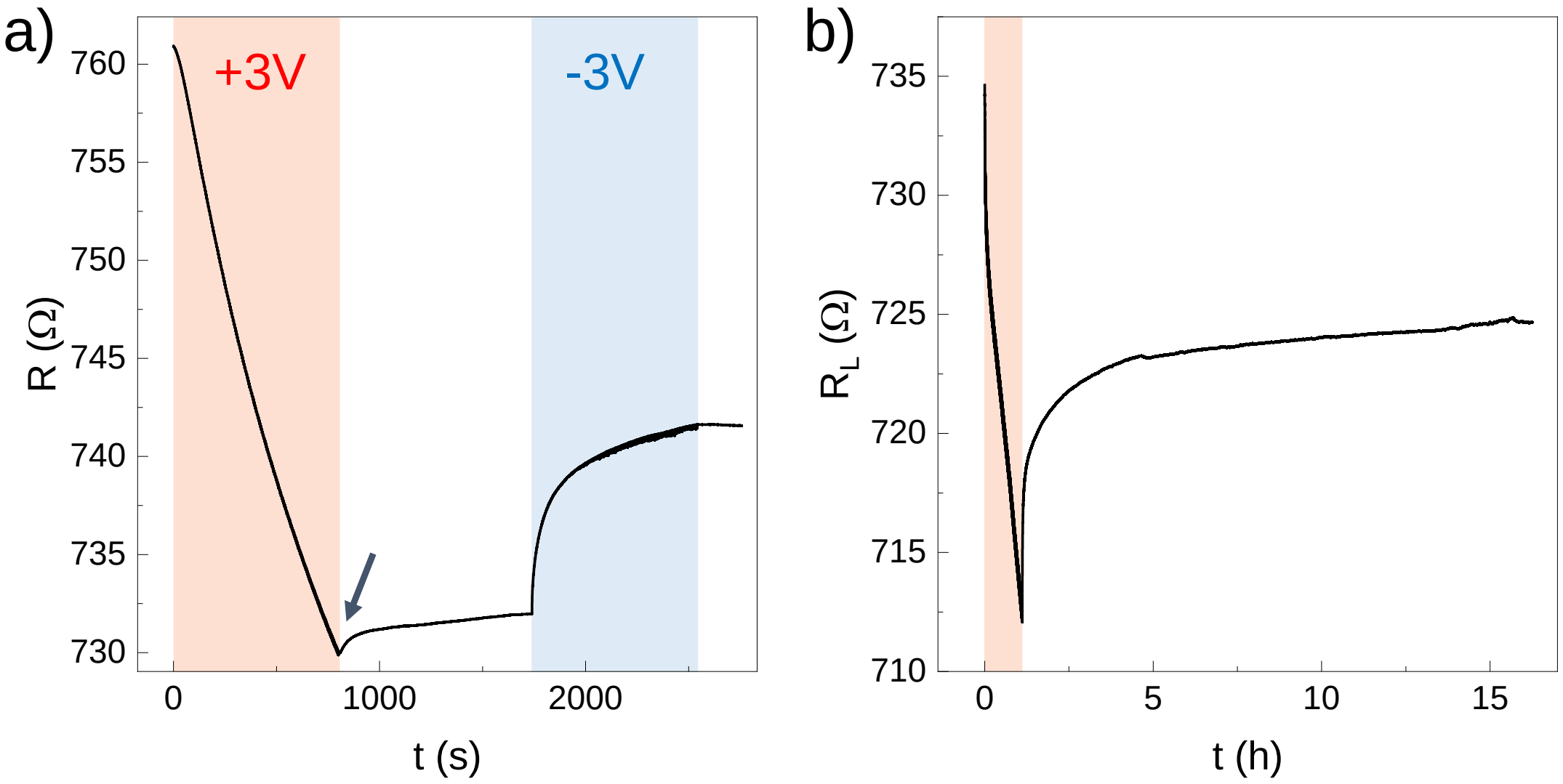

**Figure S8.1 Evolution of $R_L$ with gating**. (a) Plot of the longitudinal resistance ($R_L$) as a function of time (*t*) measured in two-pint geometry during a gating experiment. First a positive gate voltage ($V_G$=+3V red) is applied for 800s, then RL is left to stabilize for 1000s before a negative $V_G$=-3V (blue) is applied for 800s. The gray arrow indicates the initial steep increase of $R_L$ after positive gating due to the re-oxidation of Gd. (b) Plot of $R_L$ as a function of *t* measured during more than 15 h after gating.

The reversible reduction of $CuO_X$ to Cu is also reflected in the anomalous Hall effect (AHE) measurements shown in Figure S8.2a, which shows the plot of the Hall resistance as a function of out-of-plane field measured in the 3 stages of the gating experiment correspondent to Figure 5b of the main text. Initially, in the pristine state the 4-point resistance ($R_{4w}$) of the device was 83.6 Ω and $R_{AHE}$=0.45 Ω. After application of +3V during 2000s the Cu layer gets deoxidized after positive gating ($R_{4w}$=58.5 Ω)., and $R_{AHE}$ is reduced to 0.2 Ω. This is because, at equal total applied current I, the current density in the magnetic Co layer is lower when Cu is metallic. Similarly, when Cu gets partly oxidized after 1000s of gating at -3V ($R_{4w}$=68.3 Ω), $R_{AHE}$ rises back to 0.3 Ω as the current density in the magnetic layer increases.

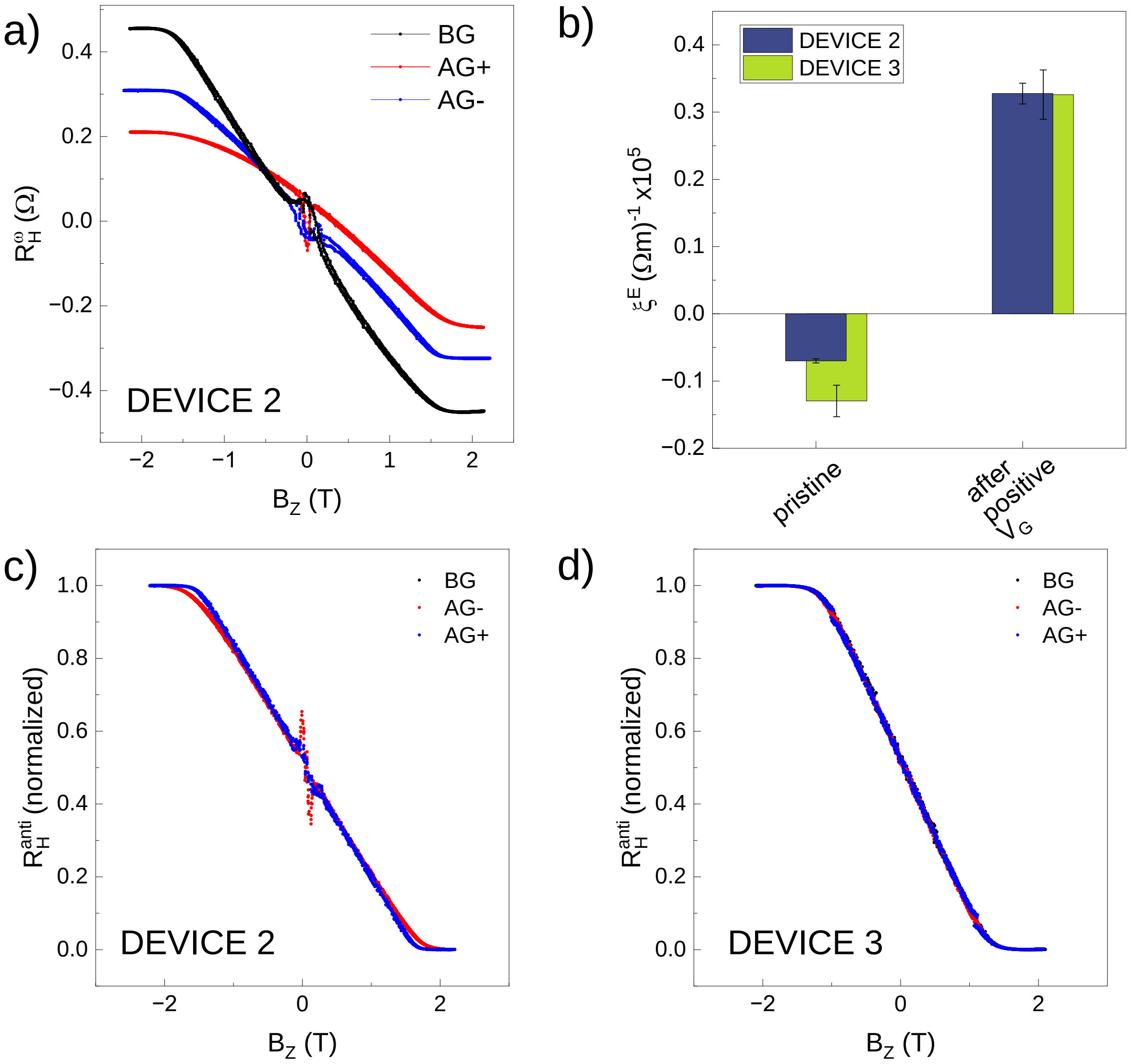

**Figure S8.2 Evolution of the AHE with gating**. a) Plot of the Hall resistance ($R_H$) as a function of out-of-plane field ($B_z$) of a Co/CuO$_X$ gated device measured before gating (BG black), after application of a positive gate (AG+ red) and after application of a negative gate (AG− blue). b) Plot of the damping-like torque efficiency ($\xi_{DL}$) measured in two different devices in the pristine state and after gating with a positive gate voltage ($V_G$). c,d) Plot of the normalized Hall resistance ($R^{anti}_{H}$) as a function of out-of-plane field ($Bz$) of the two Co/CuO$_X$ gated devices measured before gating (BG black), after application of a positive gate (AG+ red) and after application of a negative gate (AG− blue). We note that the symmetric part of the signal, mainly related to the in-plane component of the magnetization, was removed from the raw data as this contribution may depend on the initial magnetization conditions and slight field misalignment in a random manner.

The measurement of the AHE also provides information on the oxidation state of the FM. Specifically the out-of-plane saturation field ($H_S$) is proportional to $M_S$ as: $H_S = 2K_{eff}/\mu_0 M_S$, where $K_{eff}$ is the anisotropy constant that combines contributions from shape anisotropy, crystalline anisotropy, and interface effects and $\mu_0$ is the permeability of free space. $H_S$ can be determined by measurements of the Anomalous Hall Effect (AHE), as shown in Figure 1b of the main text. In all our samples $H_S$ is independent of the capping layer, but only on the thickness of the magnetic layer. Further analysis of the anomalous Hall resistance signals which are solely driven by the current through Co and provide a qualitative insight on the interfacial perpendicular magnetic anisotropy and saturation magnetization through the analysis of the out-of-plane saturation field. Minor changes in the saturation field ($H_S$) might indicate a change in the above two properties (one of them or both) linked to the Co oxidation or changes in the oxidation state at the interface. The results from two prototypical devices are shown in Figure S8.2c,d. As apparent, in the case of device 2 (same shown in Figure 5b of the main text) a small change of the shape of the AHE signal near $H_S$ can be observed after application of a positive gate voltage ($V_G$). The original $H_S$ is recovered after application of a negative gate voltage. On the other hand, for Device 3 no change is observed across the full gating cycle. Therefore, we conclude that minor oxidation of Co close to the Cu interface is possible and could occur in our experiments depending on the specific gating conditions and devices. Note that this change of the shape of the AHE loop could also be caused by a change of interface anisotropy going from a Co/$CuO_x$ to a Co/Cu interface, not necessarily indicative of the formation of $CoO_x$. However, because the torque magnitude and sign measured for the two devices of Figure RR1 are the same within the error bar (Figure S8.2b), we conclude that the presence of $CoO_x$, if any, cannot be responsible for the large amplitude and sign variation that we observe when gating, which is instead predominantly due to the oxidation of Cu. Previous literature reports [*J. Feng et al., Phys. Rev. Applied 13, 044029 (2020), S. Wu et al., Appl. Phys. Lett. 122, 122403 (2023)*] have also shown that the presence of $CoO_x$ has a rectification effect, enhancing the torque but not changing its sign. This is consistent with what measured for our reference samples where Co and Py were oxidized on purpose before the deposition of the Cu layer and some torque enhancement was observed (see Note 5).